\documentclass[12pt,a4paper]{article}
\usepackage{amsmath}
\usepackage{amsfonts}
\usepackage{setspace}
\usepackage{array}
\usepackage{hyperref}
\usepackage{cite}
\usepackage{amssymb}
\usepackage{graphicx}
\usepackage{float}

\usepackage{amsmath, graphicx}
\usepackage{dcolumn}
\usepackage{bm}
\usepackage{color}
\usepackage{epstopdf}
\usepackage{eurosym}
\usepackage{multirow}
\usepackage{wrapfig}
\usepackage{amsfonts}
\usepackage{amsmath}
\usepackage{hyperref}
\usepackage{graphicx,color}
\usepackage[english]{babel}
\usepackage{epsfig}
\usepackage{bm}
\usepackage{longtable}
\usepackage{verbatim}
\usepackage{longtable}
\usepackage{color}
\usepackage{subfigure}
\usepackage{multirow}
\usepackage{geometry}
\newcommand{\hs}{\hspace*{0.25cm}}

\newcommand{\be}{\begin{equation}}
\newcommand{\ee}{\end{equation}}
\newcommand{\bea}{\begin{eqnarray}}
\newcommand{\eea}{\end{eqnarray}}
\newcommand{\ben}{\begin{enumerate}}
\newcommand{\een}{\end{enumerate}}
\newcommand{\bde}{\begin{widetext}}
\newcommand{\ede}{\end{widetext}}
\newcommand{\nn}{\nonumber}
\newcommand{\crn}{\nonumber \\}

\newcommand{\al}{\alpha}

\newcommand{\va}{\varphi}
\newcommand{\om}{\omega}

\newcommand{\+}{\dagger}
\newcommand{\fr}{\frac}

\newcommand{\bc}{\begin{center}}
\newcommand{\ec}{\end{center}}

\newcommand{\de}{\delta}
\newcommand{\De}{\Delta}

\newcommand{\La}{\Lambda}
\newcommand{\si}{\sigma}

\newcommand{\Revised}[1]{{\color{blue}#1}}

\providecommand{\keywords}[1]{\textbf{\textit{Keywords--}} #1}
\title{$A_4$-based model with linear seesaw scheme\\
for lepton mass and mixing}

\author{V. V. Vien$^{1,2}\footnote{\footnotesize vovanvien@tdmu.edu.vn}$, H. N. Long$^{3,4}\footnote{\footnotesize hoangngoclong@tdmu.edu.vn (corresponding author)}$ \\ \small{$^1$Institute of Applied Technology, Thu Dau Mot University, }\\
\small{Binh Duong Province, Vietnam.}\\
\small{$^2$Department of Physics, Tay Nguyen University, 
Daklak, Vietnam.}\\
\small{$^3$ Center for Forecasting Studies,  Thu Dau Mot University,}\\
\small{Binh Duong Province, Vietnam}
\\
\small{$^4$Institute of Physics, Vietnam Academy of Science and Technology,}\\
\small{10 Dao Tan, Ba Dinh, Hanoi Vietnam.}}
\date{}

\begin{document}

\maketitle

\begin{abstract}
We suggest a low-scale model based on $A_4\times Z_4 \times Z_2$ symmetry and a global lepton number $U(1)_L$ symmetry capable of generating the current neutrino data. The neutrino mass smallness is reproduced by the linear seesaw mechanism. The model can explain the current observed pattern of lepton mixing in which the reactor and atmospheric angles
get the best-fit values, and the solar angle and Dirac phase lie within $3\,\si $ limits.
The obtained values of the sum of neutrino mass and the effective neutrino mass are below the present experimental limits.

\end{abstract}

\keywords{Models beyond the standard model; Neutrino mass and mixing; Non-standard-model neutrinos, right-handed neutrinos, discrete symmetries.}

\newpage
\section{\label{intro}Introduction}

\hs\hs The current neutrino data given in Ref. \cite{Salas2021} as shown in Table \ref{Salas2021} confirms that the Standard Model (SM) need to be extended.

\begin{table}[ht]
\begin{center}
\caption{\label{Salas2021} Neutrino oscillation parameters taken from Ref. \cite{Salas2021}.}
\vspace{0.25cm}
\begin{tabular}{|c|c|c|c|c|c|c|c|c|c|c|}\hline
\multirow{2}{2.3cm}{\hfill Parameters  \hfill }&\hspace{0.1 cm} $\mathrm{Normal\hspace{0.1cm} hierarchy\, (NH)}$ \hspace{0.15 cm}&\hspace{0.15 cm}$\mathrm{Inverted \hspace{0.1cm}hierarchy\, (IH)}$\hspace{0.15 cm} \\
\cline{2-3}   & $\mathrm{bfp}\pm 1\sigma \, (3\sigma \,\, \mathrm{range})$  & $\mathrm{bfp}\pm 1\sigma \, (3\sigma \,\, \mathrm{range})$ \\  \hline
$\frac{\Delta m^2_{21} (\mathrm{meV}^2)}{10}$&$7.50^{+0.22}_{-0.20}\, (6.94-8.14)$& $7.50^{+0.22}_{-0.20}\, (6.94-8.14)$\\
$\frac{|\Delta m^2_{31}| (\mathrm{meV}^2)}{10^{3}}$&$2.55^{+0.02}_{-0.03} \, (2.47-2.63)$& $2.45^{+0.02}_{-0.03}$ \, (2.37-2.53)\\
$\sin^2\theta_{12}$&\hspace{0.1cm}$0.318\pm 0.016\, (0.271-0.369)$ \hspace{0.1cm}&\hspace{0.1cm} $0.318\pm 0.016\, (0.271-0.369)$ \hspace{0.1cm}\\
$\sin^2\theta_{23}$&\hspace{0.1cm}$0.574\pm 0.014 \, (0.434-0.610)$ \hspace{0.1cm}&\hspace{0.1cm} $0.578^{+0.010}_{-0.017}\, (0.433-0.608)$ \hspace{0.1cm}\\
$\frac{\sin^2\theta_{13}}{10^{-2}}$&  $2.200^{+0.069}_{-0.062}\, (2.00-2.405)$& $2.225^{+0.064}_{-0.070}\, (2.018-2.424)$\\
$\delta/\pi$&  $1.08^{+0.13}_{-0.12}\, (0.71-1.99)$ & $1.58^{+0.15}_{-0.16}\, (1.11-1.96)$ \\
 \hline
\end{tabular}
\end{center}
\vspace{-0.25cm}
\end{table}

Although the absolute neutrino mass remains unknown, the KATRIN Collaboration has reported an upper limit of
$m_\nu < 1.1\, \mathrm{eV}$ \cite{Katrin21PRL, Katrin21PRD} or an improved limit of $m_\nu < 0.8\, \mathrm{eV}$ \cite{Katrin21arX}.

The seesaw mechanism \cite{seesaw123} is the most natural and elegant way to generate the small neutrino masses. However, the right-handed neutrinos mass scale is very high that cannot be reached by the near future experiments.
It is to be noted that, in the linear seesaw mechanism \cite{seesaw1, seesaw2, seesaw3, seesaw4, seesaw5} the neutrino mass smallness
can arise as a consequences of new physics at $\mathrm{TeV}$ scale which may be performed by the LHC experiments. In such models, both non-renormalizable and renormalizable terms are encompassed
 to accommodate the observed pattern of neutrino masses and mixings.

Discrete symmetries are useful tools for
explaining the observed fermion mass and mixing patterns in which $A_4$ symmetry has been applied in various works \cite{A41, A42, A43, A44, A46, A47, A48, A49, A410, A411, A412, A413, A414, A415, A416, A417, A418, A419, A420Ishimori10, A421VL2015, A422, A423, A424, A425}. The linear seesaw mechanism\footnote{Linear seesaw for Majorana neutrino has been discussed in Refs. \cite{LseesawA41,LseesawA45}. Here, we consider another scenarios in which the heavy neutral
singlet leptons having both left-and right-handed helicities ($N_{L, R}$ and $S_{L, R}$) are purely Dirac fermions which are similar to those of Refs. \cite{LseesawA45,LseesawA46}. However, our present model differs crucially from those of Refs. \cite{LseesawA45,LseesawA46} in that our model is based on another and smaller
abelian symmetry $Z_2$ (in stead of $Z_3$ symmetry) and with a fewer number of scalars (eight $SU(2)_L$ singlets instead of nine singles).} combined with
non-Abelian discrete symmetries has been studied in Refs. \cite{LseesawA41, LseesawS32, LseesawA43, LseesawAD274, LseesawA45, LseesawA46, LseesawD277, LseesawS48, LseesawS49}. However, the mentioned works contain non minimal scalar sectors with many $SU(2)_L$ doublets (hereafter called doublet) and they are differences with the current study. Namely, in previous works \cite{LseesawA41, LseesawS32, LseesawA43, LseesawAD274, LseesawA45, LseesawA46, LseesawD277, LseesawS48, LseesawS49}, the quark and/or lepton masses and mixings are generated by i) other non-Abelian discrete symmetries \cite{
LseesawAD274, LseesawD277, LseesawS48, LseesawS49}, ii) other Abelian discrete symmetries \cite{LseesawA41, LseesawS32, LseesawA43, LseesawAD274, LseesawA45, LseesawA46, LseesawD277, LseesawS48, LseesawS49}, iii) other gauge symmetries \cite{LseesawS32, LseesawA43, LseesawAD274, LseesawA45, LseesawA46, LseesawD277, LseesawS48, LseesawS49}, iv) and many scalar fields:
up to nine doublets \cite{LseesawA41}, three triplets and fifteen singlets \cite{LseesawS32}, one doublets and ten singlets \cite{LseesawA43}, thirteen doublets and fourteen singlets \cite{LseesawAD274}, one doublets and nine singlets \cite{LseesawA45, LseesawA46}, five doublets and fifteen singlets \cite{LseesawD277}, three doublets and nine singlets \cite{LseesawS48}, and three triplets, two doublets and fifteen singlets \cite{LseesawS49}.
Therefore, it is necessary and important to search for another extension with a simpler scalar sector
for explaining
the observed neutrino data.
In this study, we suggest a SM extension based on flavor symmetry $A_4\times Z_4 \times Z_2$ with only one doublet and eight extra singlet scalars.

\section{\label{model}The model}
Besides the SM gauge symmetry, our model has been supplemented by three discrete symmetries $A_4, Z_4$ and $Z_2$, i.e., the full symmetry is $SU(2)_L\times U(1)_Y\times U(1)_L\times A_4\times  Z_4\times  Z_2 \equiv \mathbf{G}$. On the other hand, three right-handed neutrinos ($\nu_{R}$) and two types of neutral singlet leptons with two helicity states $(N_{L, R},\, S_{L, R})$ together with eight singlet scalars are additionally introduced.
Three left-handed leptons and three right-handed neutrinos together with extra neutral leptons\footnote{Each of extra neutral leptons has a lepton number of 1.} $N_{L, R}, S_{L, R}$ are put in $A_4$ triplets while three right-handed charged leptons $l_{1,2,3 R}$ are in $\underline{1}, \underline{1}'$ and $\underline{1}''$ under $A_4$ symmetry, respectively.
The assignment of leptons and scalars is given in Table \ref{lepcont}.
\begin{table}[ht]
\caption{\label{lepcont} The assignment under $\mathbf{G}$ symmetry for leptons and scalars.}
\vspace{0.25cm}
\centerline{\begin{tabular}{ccccccccccccccc|c|c|c|c|c|c|c|c|c|c|c|c|c|}
\hline
Fields& $\psi_{L}$ & $l_{1, 2, 3R}$ &$\nu_{R}$& $N_L, S_L$ &$N_{R}, S_R$ &$H$& $\phi, \varphi$ &$\chi, \rho$ \\ \hline
 $[SU(2)_L, \mathrm{U}(1)_Y]$ & $\left[2, -\frac{1}{2}\right]$ & [1, -1] &[1, 0] &[1, 0] &[1, 0] &$\left[2, \frac{1}{2}\right]$& [1, 0] & [1, 0]  \\ 
$\mathrm{U}(1)_{L}$ & $1$ &$1$&$1$&$1$&$1$&$0$& $0$ & $0$ \\ 
$A_4$ & $\underline{3}$ & $\underline{1}, \underline{1}', \underline{1}''$& $\underline{3}$  & $\underline{3}$ &$\underline{3} $ & $\underline{1}$ &$\underline{3}$ &$\underline{1}$ \\ 
$Z_4$ & $i$&$i$&$1$ &$-i$& $i$  & $1$ &$1, -1$ &$-1, -i$\\ 
$Z_2$ &$ +$ &$ -$ & $-$  & $-$&$+$ &$+$&$-$&$-, +$ \\\hline
\end{tabular}}
\end{table}

It is noted that, with the field content in Table \ref{lepcont}, in general, the basis of linear  seesaw mechanism could be taken by $(\nu_L, \nu_R^c,  N_L, N_R^C, S_L, S_R^C)$, i.e., the mass Lagrangian for neutrino is written in the form:
\bea
 -\mathcal{L}^{\mathrm{mass}}_\nu = \frac{1}{2}\left(\overline{n}_L \hs\hs \overline{n^C_R}\right)\left(%
 \begin{array}{ccc}
 m_{L} & m_{D} \\
 m_{D}^T & m_R \\
\end{array}%
\right)\left(%
\begin{array}{c}
n^C_L \\
n_R\\
\end{array}%
\right), \label{masslagrangian}
\eea
where
\bea
&&n=\left(\nu \hs\hs  N \hs\hs S\right)^T,\hs m_L=\left(%
\begin{array}{ccc}
m^L_{\nu\nu}&\,\,\, m^L_{\nu N} &\,\,\, M^L_{\nu S} \\
 m'^{L}_{\nu N} & M^L_{NN} &\,\,\, M^L_{NS} \\
 M'^L_{\nu S} & \,\,\, M'^L_{NS}  &M^L_{SS}\\
\end{array}%
\right), \crn
&&m_D=\left(%
\begin{array}{ccc}
m_{\nu\nu}&\,\,\, m_{\nu N} &\,\,\, M_{\nu S} \\
 m'_{\nu N} & M_{NN} &\,\,\, M_{NS} \\
 M'_{\nu S} & \,\,\, M'_{NS}  &M_{SS}\\
\end{array}%
\right), \hs m_R=\left(%
\begin{array}{ccc}
m^R_{\nu\nu}&\,\,\, m^R_{\nu N} &\,\,\, M^R_{\nu S} \\
 m'^R_{\nu N} & M^R_{NN} &\,\,\, M^R_{NS} \\
 M'^R_{\nu S} & \,\,\, M'^R_{NS}  &M^R_{SS}\\
\end{array}%
\right). \label{massmatrices}
\eea
In the model under consideration, upto five-dimension, the additional symmetries $A_4$, $U(1)_L$, $ Z_4$ and $Z_2$ make $m_L=0,\, m_R=0$ and $m_{\nu\nu}=0$, i.e., the mass Lagrangian for neutrino in Eqs. (\ref{masslagrangian}) and (\ref{massmatrices}) becomes:
\bea
 -\mathcal{L}^{\mathrm{mass}}_\nu&=&\frac{1}{2}\left(\bar{\nu}_L \hs\hs \bar{N}_L \hs\hs \bar{S}_L\right)\left(%
\begin{array}{ccc}
 0 &\,\,\, m_{\nu N} &\,\,\, M_{\nu S} \\
 m'_{\nu N} & M_{NN} &\,\,\, M_{NS} \\
 M'_{\nu S} & \,\,\, M'_{NS}  &M_{SS}\\
\end{array}%
\right)\left(%
\begin{array}{c}
  \nu_{R} \\
  N_{R} \\
  S_{R} \\
\end{array}%
\right)+h.c, \crn
&\equiv &\frac{1}{2} \mathrm{M}_{\mathrm{eff}}\bar{n}_L n_R + h.c\Revised{.}
\eea
The additional symmetries $U(1)_L, Z_4$ and $Z_2$ play important roles in preventing unwanted Yukawa couplings to get the desired structure of the mass matrices which are listed in Tables \ref{U1Lprevent}, \ref{Z4prevent} and \ref{Z2prevent}, respectively.

The lepton Yukawa terms, up to five-dimensions which are invariant under $\mathbf{G}$ are:
\bea
-\mathcal{L}_l&=&\fr{h_{e}}{\La }(\bar{\psi}_L\phi)_{\underline{1}} (H l_{1R})_{\underline{1}} +\fr{h_{\mu}}{\La }(\bar{\psi}_L \phi)_{\underline{1}''}(H l_{2R})_{\underline{1}'}+\fr{h_{\tau}}{\La }(\bar{\psi}_L \phi)_{\underline{1}'}(H l_{3R})_{\underline{1}''}\crn
  &+&x_{1\nu} (\bar{\psi}_L N_R)_1  \widetilde{H} +x_{2\nu} (\bar{\psi}_L S_R)_1 \widetilde{H}
  +x_{3\nu} (\overline{N}_L \nu_R)_1  \rho+x_{4\nu} (\overline{S}_L \nu_R)_1  \rho \crn
 &+& y_{1\nu} (\overline{S}_L N_R)_1 \chi  +y_{2\nu} (\overline{S}_L N_R)_{3_s}\varphi +y_{3\nu} (\overline{S}_L N_R)_{3_a}\varphi \crn
  &+&z_{1\nu} (\overline{N}_L S_R)_1 \chi +z_{2\nu}(\overline{N}_L S_R)_{3_s}\varphi + z_{3\nu} (\overline{N}_L S_R)_{3_a}\varphi+ \crn
 &+&t_{1\nu}(\overline{N}_L N_R)_1 \chi + t_{2\nu}(\overline{N}_L N_R)_{3_s}\varphi  + t_{3\nu} (\overline{N}_L N_R)_{3_a}\varphi \crn
 &+&u_{1\nu}(\overline{S}_L S_R)_1 \chi + u_{2\nu}(\overline{S}_L S_R)_{3_s}\varphi + u_{3\nu} (\overline{S}_L S_R)_{3_a}\varphi
+ \mathrm{h.c}., \label{Ylep}\eea
where $\Lambda$ is the cut-off scale, and $h_{e, \mu,\tau}, y_{i\nu}, z_{i\nu}, t_{i\nu}, u_{i\nu}$ and $x_{j \nu}\,\, (i=1,2,3; j=1,2,..,4)$ are the Yukawa-like dimensionless couplings.

There exist seven six-dimensional terms including
$\fr{1}{\La^{2}}(\overline{\psi}_L \nu_R)_{\underline{3}_{s, a}} (\widetilde{H} \varphi \rho)_{\underline{3}}$, $\fr{1}{\La^{2}}(\overline{\psi}_L \nu_R)_{\underline{1}} (\widetilde{H} \chi \rho)_{\underline{1}}$, $\fr{1}{\La^{2}}(\overline{\psi}_L N_R)_{\underline{3}_{s, a}} (\widetilde{H} \varphi \chi)_{\underline{3}}$ and $\fr{1}{\La^{2}}(\overline{\psi}_L S_R)_{\underline{3}_{s, a}} (\widetilde{H} \varphi \chi)_{\underline{3}}$, in which due to the fact that $v_H \ll
v_\phi \sim v_\varphi  \sim v_\chi \sim v_\rho \ll \Lambda$, the first three terms $\fr{1}{\La^{2}}(\overline{\psi}_L \nu_R)_{\underline{3}_{s, a}} (\widetilde{H} \varphi \rho)_{\underline{3}}$ and $\fr{1}{\La^{2}}(\overline{\psi}_L \nu_R)_{\underline{1}} (\widetilde{H} \chi \rho)_{\underline{1}}$, contribute to the left-handed neutrino mass $m_{\nu\nu}$ generated via Type II seesaw mechanism, is very small
 compared to the one generated via the canonical type-I seesaw mechanism as
 in Eq. (\ref{mnu1}) below; the next two terms $\fr{1}{\La^{2}}(\overline{\psi}_L N_R)_{\underline{3}_{s, a}} (\widetilde{H} \varphi \chi)_{\underline{3}}$ contribute to the element $m_{\nu N}$ but it is very small compared to the contribution from $(\bar{\psi}_L N_R)_1  \widetilde{H}$, and the last two terms $\fr{1}{\La^{2}}(\overline{\psi}_L S_R)_{\underline{3}_{s, a}} (\widetilde{H} \varphi \chi)_{\underline{3}}$ contribute to the element $M_{\nu S}$ but it is very small compared to the contribution from $(\bar{\psi}_L S_R)_1  \widetilde{H}$. Therefore these six-dimensional terms are heavily suppressed and we have
not encompassed them in the Lagrangian (\ref{Ylep}).\\
The vacuum expectation value (VEV) configuration of the scalars, which comes from the minimum condition of the model scalar
 potential (see, for example, Refs. \cite{scalarpoten1, scalarpoten2, scalarpoten3, scalarpoten4} for a similar and detailed analysis),
 reads:
 \bea
&&\langle H \rangle = \left( 0  \hspace{0.25cm} v_H  \right)^T, \hspace{0.15cm} \langle \phi \rangle = (\langle \phi_1 \rangle, \hspace{0.25cm} \langle \phi_2 \rangle,\hspace{0.25cm} \langle \phi_3 \rangle), \hspace{0.15cm} \langle \phi_1 \rangle =\langle \phi_2 \rangle=\langle \phi_3 \rangle= v_{\phi},  \crn
&& \langle \varphi \rangle = (0,\hspace{0.275cm} \langle \varphi_2 \rangle,\hspace{0.275cm} 0 \rangle), \hspace{0.15cm} \langle \varphi_2 \rangle= v_{\varphi}, \hspace{0.15cm} \langle \chi \rangle= v_{\chi}, \hspace{0.25cm} \langle \rho \rangle = v_\rho. \label{scalarvev}
\eea
It is noted that, with the VEV
alignment in Eq. (\ref{scalarvev}), $\phi$ breaks $A_4$ down to
$Z_3$ while $\varphi$ breaks $A_4$ down to
$Z_2$ symmetry.
The electroweak symmetry is broken at a low scale, $v_H = 246 \, \mathrm{GeV}$ (see below). In this work, we assume that the VEV of singlets and the cut-off scale are at a very high scale,
\bea
v_\phi \simeq  v_\varphi  \simeq v_\chi \simeq v_\rho \simeq 10^{10} \, \mathrm{GeV}, \hs \Lambda \simeq 10^{13}\, \mathrm{GeV}. \label{vevscales}
\eea
As shown in Appendix \ref{anomaly}, in the case of the quark fields, under $\mathbf{G}$ symmetry, they transforms as $Q_{1L} =\left(u_{1L} \hs d_{1L}\right)^T\sim \left(2, \frac{1}{6}, -\frac{1}{3}, \underline{1}', 1, +\right)$, $Q_{2L} =\left(u_{2L} \hs
d_{2L}\right)^T\sim \left(2, \frac{1}{6}, -\frac{1}{3}, \underline{1}^{''}, 1, +\right)$, $Q_{3L} =\left(u_{3L} \hs
d_{3L}\right)^T\sim \left(2, \frac{1}{6}, -\frac{1}{3}, \underline{1}, 1, +\right)$, $u_{R} \sim \left(1, \frac{2}{3}, -\frac{1}{3}, \underline{3}, 1, -\right)$ and $d_{R} \sim \left(1, -\frac{1}{3}, -\frac{1}{3}, \underline{3}, 1, -\right)$, all anomalies are canceled within each generation. On the other hand, the SM quark masses are generated by the following Yukawa terms:
\bea
-\mathcal{L}_q&=&\fr{h^u_{1}}{\La}(\bar{Q}_{1L}\widetilde{H})_{\underline{1}'} (\phi u_{R})_{\underline{1}^{''}}
+\fr{h^u_{2}}{\La}(\bar{Q}_{2L}\widetilde{H})_{\underline{1}^{''}} (\phi u_{R})_{\underline{1}'}
+\fr{h^u_{3}}{\La}(\bar{Q}_{3L}\widetilde{H})_{\underline{1}} (\phi u_{R})_{\underline{1}}\crn
&+&\fr{h^d_{1}}{\La}(\bar{Q}_{1L}H)_{\underline{1}'} (\phi d_{R})_{\underline{1}^{''}}
+\fr{h^d_{2}}{\La}(\bar{Q}_{2L}H)_{\underline{1}^{''}} (\phi d_{R})_{\underline{1}'}
+\fr{h^d_{3}}{\La}(\bar{Q}_{3L}H)_{\underline{1}} (\phi d_{R})_{\underline{1}}+ \mathrm{h.c}. \label{Yquark}\eea
However, in this study, we only concentrate on the lepton sector without mentioning the quark one.

Looking at  the scalar content of the model, it is realized that only fields in the doublet $H$ are complex, while all  remaining scalar singlets $\phi, \va , \rho$ and $\chi$ are real fields. In addition, the masses of gauge bosons are due to just the doublet $H$. Hence, we can express the scalar fields as follows
\bea
&&H=
\left(%
\begin{array}{c}
G_W^+ \\
\fr 1{\sqrt{2}} \left(v_{W} + R_H + i G_Z\right)\\
\end{array}%
\right), \hs  \rho = v_\rho + R_\rho, \crn
&&\phi_i = v_{\phi_i} + R_{\phi_i}, \hs\hs\,\, \va_i  =  v_{\va_i} + R_{\va_i}, \hs\hs\,\, \chi =  v_\chi + R_\chi, \label{higg1}\\
&& v_{\phi_i}  =  v_\phi \,, i= 1,2,3\, ,\hs   v_{\va_2} = v_\va \, ,\hs  v_{\va_n} = 0\,, n = 1, 3,\crn
&&  v_W  =  v_H = 246 \, \textrm{GeV}, \label{higgm}
\eea
where $G_W^+$ and $ G_Z$ are Goldstone bosons eaten by $W^+$ and $Z$ bosons, respectively.
The dark matter candidate is the {\it lightest} state which is combination of CP-even components
the $Z_2$ {\it odd} scalars $ R_\phi, R_\va , R_\chi$. A detailed
study of this issue is beyond the scope of this study and would necessitate a further study.
\section{\label{lepton}Lepton masses and mixings}
Using the tensor product of $A_4$ group \cite{A420Ishimori10}, after symmetry breaking, i.e., the scalar fields $H$ and $\phi$
get their VEVs  given in Eq. (\ref{scalarvev}), we get the charged lepton mass matrix 
\bea
M_l=\fr{v_H v_\phi}{\Lambda}\left(
\begin{array}{ccc}
  h_e &\,\,\, h_\mu &\,\,\, h_\tau \\
  h_e &\,\,\, \om^2 h_\mu  &\,\,\, \om h_\tau \\
  h_e &\,\,\, \om h_\mu &\,\,\, \om^2 h_\tau \\
\end{array}%
\right),
\eea
which can be diagonalised by $U_{l,r}$ given by
\bea
&&U^\+_l=\fr{1}{\sqrt{3}}\left(%
\begin{array}{ccc}
  1 &\,\,\, 1 &\,\,\, 1 \\
  1 &\,\,\, \om &\,\,\, \om^2 \\
  1 &\,\,\, \om^2 &\,\,\, \om \\
\end{array}%
\right),\hs U_r=\mathbf{I}_{3\times 3}, \hs \left(\om=e^{i\frac{2\pi}{3}}\right), \label{Uclep}\\
&&m_e =\sqrt{3} h_e v_H \frac{ v_{\phi}}{\Lambda},\hs  m_{\mu}=\sqrt{3} h_\mu v_H \frac{ v_{\phi}}{\Lambda},\hs
m_{\tau}=\sqrt{3} h_\tau v_H \frac{ v_{\phi}}{\Lambda}.\label{memtv}\eea
The left-handed charged-lepton mixing matrix $U_l$ in Eq. (\ref{Uclep}) is non trivial, and then it will contribute to 
the lepton mixing matrix.

Now, using the expansion of $\phi_i$ and $H$, $\phi_i=\langle \phi_i\rangle +\Phi_i$ and $H=(H^+\hs H^0)^T$, Eq. (\ref{Ylep}) yields the lepton flavor changing interactions:
\bea
-\mathcal{L}_{clep}&\supset& \fr{h_e v_{\phi}}{\La }\left[(\bar{\nu}_{2L}+\bar{\nu}_{3L})H^+ +(\bar{l}_{2L}+\bar{l}_{3L})H^0\right] l_{1R} \crn
&+&\fr{h_\mu v_{\phi}}{\La}\left[(\bar{\nu}_{1L}+\om \bar{\nu}_{3L})H^+ +(\bar{l}_{1L}+\om \bar{l}_{3L})H^0\right] l_{2R} \crn
&+&\fr{h_\tau v_{\phi}}{\La}\left[(\bar{\nu}_{1L}+\om \bar{\nu}_{2L})H^+ +(\bar{l}_{1L}+\om \bar{l}_{2L})H^0\right] l_{3R} + H.c.
 \label{CLV}
 \eea
Equation (\ref{CLV}) implies that the usual Yukawa couplings are proportional to
$\frac{v_{\phi}}{\La}$ and the lepton flavor changing processes are suppressed by the factor $\fr{v_{\phi}}{\La}\fr{1}{G_F^2M_H^2}$ where $G_F=\frac{g^2}{4\sqrt{2}m^2_W}$ and $M_H$ is the mass scale of the heavy scalars. For further details, the reader is referred to Refs. \cite{br1,br2,br3,br4}.

Furthermore, Eq. (\ref{memtv}) implies that our model can successfully accommodate the SM charged lepton masses. Indeed, comparing the model result in Eq. (\ref{memtv}) with the experimental values \cite{PDG2020},  $m_e=0.51099 \,\mathrm{MeV},  m_\mu = 105.65837\,\mathrm{MeV}, m_\tau = 1776.86 \,\mathrm{MeV}$, and taking the values in Eq. (\ref{vevscales}), we get:
\bea
h_{e} \sim 10^{-3},\hs h_{\mu} \sim 10^{-1},\hs h_{\tau}\sim 1.0.
\eea
Now we turn to the neutrino sector. From Eq. (\ref{Ylep}), after symmetry breaking, 
we obtain the following neutrino mass matrices:
 \bea
&&m_{\nu N} = x_{1\nu} v_H \textbf{I}\equiv a_1 \textbf{I}, \hs M_{\nu S} = x_{2\nu} v_H \textbf{I}\equiv a_2 \textbf{I}, \label{submatrix1}\\
&&m'_{\nu N}=  x_3  v_\rho \textbf{I}\equiv a_3 \textbf{I}, \hs\hs  M'_{\nu S} = x_4  v_\rho \textbf{I}\equiv a_4 \textbf{I}, \label{submatrix2} \\
&&M'_{NS}= \left(%
\begin{array}{ccc}
  y_{1\nu} v_\chi & 0  & (y_{2\nu}-y_{3\nu}) v_\varphi  \\
  0 & y_{1\nu} v_\chi &0 \\
  (y_{2\nu}+y_{3\nu}) v_\varphi  & 0& y_{1\nu} v_\chi \\
\end{array}%
\right)\equiv  \left(%
\begin{array}{ccc}
  b_1 & 0 & b_2 - b_3 \\
  0 & b_1 & 0 \\
  b_2 + b_3 & 0  & b_1 \\
\end{array}%
\right), \label{submatrix3}\\
&&M_{NS} =\left(%
\begin{array}{ccc}
z_{1\nu} v_\chi & 0  & (z_{2\nu}-z_{3\nu}) v_\varphi  \\
0 & z_{1\nu} v_\chi &0 \\
(z_{2\nu}+z_{3\nu}) v_\varphi  & 0& z_{1\nu} v_\chi \\
\end{array}%
\right)\equiv  \left(%
\begin{array}{ccc}
c_1 & 0 & c_2 - c_3 \\
0 & c_1 & 0 \\
c_2 + c_3 & 0  & c_1 \\
\end{array}%
\right),\hs\hs \label{submatrix4}\eea
\bea
&&M_{NN}=\left(%
\begin{array}{ccc}
t_{1\nu} v_\chi & 0  & (t_{2\nu}-t_{3\nu}) v_\varphi  \\
0 & t_{1\nu} v_\chi &0 \\
(t_{2\nu}+t_{3\nu}) v_\varphi  & 0& t_{1\nu} v_\chi \\
\end{array}%
\right)\equiv  \left(%
\begin{array}{ccc}
d_1 & 0 & d_2 - d_3 \\
0 & d_1 & 0 \\
d_2 + d_3 & 0  & d_1 \\
\end{array}%
\right), \label{submatrix5}\\
&&M_{SS}=\left(%
\begin{array}{ccc}
u_{1\nu} v_\chi & 0  & (u_{2\nu}-u_{3\nu}) v_\varphi  \\
0 & u_{1\nu} v_\chi &0 \\
(u_{2\nu}+u_{3\nu}) v_\varphi  & 0& u_{1\nu} v_\chi \\
\end{array}%
\right)\equiv  \left(%
\begin{array}{ccc}
g_1 & 0 & g_2 - g_3 \\
0 & g_1 & 0 \\
g_2 + g_3 & 0  & g_1 \\
\end{array}%
\right)\hspace{-0.1 cm}.\hspace{0.7 cm}\label{submatrix6}\eea
The effective neutrino mass matrix, 
in the basis ($\nu$ , N, S), takes the form
  \bea
 M_{\mathrm{eff}}&=& 
\left(%
 \begin{array}{ccc}
  0&M_D \\
 M^T_D & M_R \\
\end{array}%
\right),  \label{Meff0}\eea
where we have introduced the following matrices
\bea
M_D&=&(m_{\nu N} \,\,\, M_{\nu S}),\,\, M^T_D= \left(%
 \begin{array}{c}
  m'_{\nu N}\\
 M'_{\nu S} \\
\end{array}%
\right), \,\, M_R= \left(%
 \begin{array}{cc}
M_{NN} & M_{NS} \\
M'_{NS}  & M_{SS} \\
\end{array}%
\right), \nn
   \eea
and all the entries of $M_{\mathrm{eff}}$ are identified in Eqs. (\ref{submatrix1}) - (\ref{submatrix6}).
The active Dirac neutrino mass matrix, $m_\nu=- M_D M^{-1}_R M^T_D$, then gets the following form:
   \bea
   m_\nu &=& \left(%
\begin{array}{ccc}
 A_1&0 &A_4\\
 0&A_2&0  \\
 A_5 &0 &A_3 \\
\end{array}%
\right),  \label{mnu1}
\eea
where
\bea
A_i&=&- \alpha_i a_1 + \beta_i a_2\,\, (i=1,2,3),\crn
A_4&=& \alpha_{13} a_1 a_3 +\alpha_{23}  a_2 a_3+ \alpha_{14} a_1 a_4+ \alpha_{24}a_2 a_4, \crn
A_5&=&\beta_{13} a_1 a_3+\beta_{23}  a_2 a_3+\beta_{14}  a_1 a_4+ \beta_{24} a_2 a_4,  \label{A12345}
\eea
with $\al_{k}, \beta_{k}, \al_{pq}$ and $\beta_{pq} \, (k=1,2,3; pq=13, 23, 14, 24)$ are explicitly identified in Appendices \ref{Appenalbeta} and \ref{Appenalbetaij}.

In order to diagonalise the mass matrix $m_{\nu}$ in Eq. (\ref{mnu1}), we define a Hermitian matrix $\mathcal{M}^2_\nu$, given by
\bea
\mathcal{M}^2_\nu&=&m_{\nu} m^\+_{\nu}
=
 \left(
\begin{array}{ccc}
 a_0^2 & 0 & d_0 e^{-i \psi} \\
 0 & b_0^2 & 0 \\
 d_0 e^{i \psi} & 0 & c_0^2 \\
\end{array}
\right),\label{Mnusq}
\eea
where
\bea
&&a_0^2 = A_{01}^2 + A_{04}^2, \hs
b_0^2 = A_{02}^2, \hs
c_0^2 = A_{03}^2 + A_{05}^2, \crn
&&d_0 = \sqrt{
  A_{03}^2 A_{04}^2 + A_{01}^2 A_{05}^2 +
   2 A_{01} A_{03} A_{04} A_{05} \cos(\psi_1 + \psi_3 - \psi_4 - \psi_5)}, \\
&&\psi=\arctan \left(\frac{A_{03} A_{04} \sin (\psi_{3}-\psi_{4})-A_{01} A_{05} \sin (\psi_{1}-\psi_{5})}{A_{01} A_{05} \cos (\psi_{1}-\psi_{5})+ A_{03} A_{04} \cos (\psi_{3}-\psi_{4})}\right),  \label{psi}
\eea
and $\psi_i =\mathrm{arg}(A_{i})$, $A_{0i}=|A_{i}|\,\, (i=1,2,..,5)$ are real and positive parameters.

The matrix $\mathcal{M}^2_\nu$ in Eq.(\ref{Mnusq}) is diagonalised by the unitary matrix $U_{\nu}$, satisfying
 \bea
U_{\nu}^\+ \mathcal{M}^2_\nu U_{\nu}=\left\{
\begin{array}{l}
\left(
\begin{array}{ccc}
m^2_1 & 0 & 0 \\
0 & m^2_{2} & 0 \\
0 & 0 & m^2_{3}%
\end{array}%
\right),\hspace{0.25cm} U_{\nu}=\left(
\begin{array}{ccc}
\cos\theta_\nu & 0 & \sin \theta_\nu. e^{-i \psi}\\
 0 & 1 & 0 \\
 -\sin \theta_\nu. e^{i \psi} & 0 &\cos\theta_\nu  \\
\end{array}
\right) \hspace{0.2cm}\mbox{for NH,}\ \  \\
\left(
\begin{array}{ccc}
m^2_{3} & 0 & 0 \\
0 & m^2_{2} & 0 \\
0 & 0 & m^2_1
\end{array}%
\right) ,\hspace{0.2cm} U_{\nu}=\left(
\begin{array}{ccc}
 \sin \theta_\nu & 0 &  -\cos\theta_\nu. e^{-i \psi}\\
 0 & 1 & 0 \\
 \cos\theta_\nu. e^{i \psi} & 0 & \sin \theta_\nu \\
\end{array}
\right) \hspace{0.2cm}\mbox{for IH, }%
\end{array}%
\right.  \label{Unuv}
\eea
where
\bea
&&m^2_{1,3} =\frac{1}{2} \left(a_0^2 + c_0^2 \mp \sqrt{\left(a_0^2 - c_0^2\right)^2 + 4 d_0^2}\right)\equiv \Gamma^2_1\mp \Gamma^2_2, \,\,\, m^2_2=b_0^2, \label{m123sq}\\
&&\theta_\nu=\arctan\left(\frac{d_0}{m^2_{1}-c_0^2}\right)=\arctan\left(\frac{d_0}{a_0^2-m^2_{3}}\right).  \label{theta}
\eea
The sign of $\Delta  m^2_{31}$ plays an important role in determining the hierarchy of neutrino mass spectrum
 where $m_1< m_{2}< m_{3}$ for NH and $m_{3}< m_1< m_{2}$ for IH. It is
 noted that the eigenvalue $m_2^2=A^2_{02}$ corresponds to the second
 neutrino eigenvector $\varphi_2=(0\hs 1\hs 0)^T$. Thus, the neutrino mass hierarchy
 should be either $\left(m^2_1,\, m^2_2,\, m^2_3\right)$ or $\left(m^2_3,\, m^2_2,\, m^2_1\right)$. As will
  see below, both these two cases are inconsistent with the observed
  neutrino oscillation data, i.e., the model under consideration can predict both the normal
   and inverted ordering of the active neutrino masses. The eigenvalues and corresponding
   vectors of $ \mathcal{M}^2_\nu$ in Eq. (\ref{Mnusq}), for the two mass hierarchies, are defined by:
\be
U_{\nu }^\+ \mathcal{M}^2_\nu U_{\nu }=\left\{
\begin{array}{l}
\left(
\begin{array}{ccc}
m^2_1 & 0 & 0 \\
0 & m^2_{2} & 0 \\
0 & 0 & m^2_{3}%
\end{array}%
\right) ,\hspace{0.1cm} U_{\nu }=\left(
\begin{array}{ccc}
\cos \theta_\nu & 0 & -\sin\theta_\nu e^{-i \psi}\\
 0 & 1 & 0 \\
\sin\theta_\nu e^{i \psi} & 0 & \cos \theta_\nu \\
\end{array}
\right) \hspace{0.2cm}\mbox{for NH,}\ \  \\
\left(
\begin{array}{ccc}
m^2_{3} & 0 & 0 \\
0 & m^2_{2} & 0 \\
0 & 0 & m^2_1
\end{array}%
\right) ,\hspace{0.1cm} U_{\nu }=\left(
\begin{array}{ccc}
 \sin\theta_\nu & 0 &\cos \theta_\nu e^{-i \psi}\\
 0 & 1 & 0 \\
 -\cos \theta_\nu e^{i \psi} & 0 &\sin\theta_\nu  \\
\end{array}
\right) \hspace{0.2cm}\mbox{for IH,}%
\end{array}%
\right.  \label{Unu}
\ee
where $\psi$, $m^2_{1, 2,3}$ and $\theta_\nu$ are given in Eqs. (\ref{psi}), (\ref{m123sq}) and (\ref{theta}), respectively.

The leptonic mixing, $U_{\mathrm{lep}}=U_{l}^{\dag} U_{\nu }$, obtained from Eqs. (\ref{Uclep}) and (\ref{Unu}):
\bea
U_{\mathrm{lep}}=
\left\{
\begin{array}{l}
\hspace{-0.15 cm}\fr{1}{\sqrt{3}}\left( \begin{array}{ccc}
\hspace{-0.1 cm}\cos\theta_\nu + \sin\theta_\nu. e^{i \psi }      &1 & \cos\theta_\nu - \sin\theta_\nu. e^{-i \psi}  \\
\hspace{-0.1 cm}\cos\theta_\nu + \om^2 \sin\theta_\nu. e^{i \psi}\hs &\om &\hs \om^2\cos\theta_\nu - \sin\theta_\nu. e^{-i \psi}  \\
\hspace{-0.1 cm}\cos\theta_\nu + \om \sin\theta_\nu. e^{i \psi}&\om^2  & \om \cos\theta_\nu - \sin\theta_\nu. e^{-i \psi} \\
\end{array}\hspace{-0.1 cm}\right) \hspace{0.1cm}\mbox{for NH},  \label{Ulep}  \\
\hspace{-0.15 cm}\fr{1}{\sqrt{3}}\left( \begin{array}{ccc}
\hspace{-0.1 cm}\sin\theta_\nu-\cos\theta_\nu. e^{i \psi }      &1 & \sin\theta_\nu+\cos\theta_\nu. e^{-i \psi}  \\
\hspace{-0.1 cm}\sin\theta_\nu - \om^2 \cos\theta_\nu. e^{i \psi}\hs &\om &\hs \om^2\sin\theta_\nu +\cos\theta_\nu. e^{-i \psi}  \\
\hspace{-0.1 cm}\sin\theta_\nu - \om \cos\theta_\nu. e^{i \psi}&\om^2  & \om \sin\theta_\nu+\cos\theta_\nu. e^{-i \psi} \\
\end{array}\hspace{-0.1 cm}\right) \hspace{0.1cm}\mbox{for IH}.
\end{array}%
\right.
\eea

In the three-neutrino framework, the lepton mixing matrix defined in Eq. (\ref{Ulep}) and with the observed neutrino mixing angles given in Table \ref{Salas2021} satisfy the following relations
\bea
&& s_{13}^2=\left| U_{1 3}\right|^2=\left\{
\begin{array}{l}
\frac{1}{3} \left(1 - \sin 2\theta_\nu  \cos\psi\right) \hspace{0.2cm}\mbox{for  NH},    \\
\frac{1}{3} \left(1 + \sin 2\theta_\nu  \cos\psi\right)  \hspace{0.2cm}\,\mbox{for  IH},
\end{array}%
\right. \label{s13sq}\\
&&t_{12}^2 = \left|\fr{U_{1 2}}{U_{1 1}}\right|^2=\left\{
\begin{array}{l}
\frac{1}{1 + \sin 2\theta_\nu  \cos\psi} \hspace{0.2cm}\mbox{for  NH},    \\
\frac{1}{1 - \sin 2\theta_\nu  \cos\psi}  \hspace{0.2cm}\mbox{for  IH},
\end{array}%
\right. \label{t12sq}\\
&&t_{23}^2=\left|\fr{U_{2 3}}{U_{3 3}}\right|^2=\left\{
\begin{array}{l}
\frac{\sqrt{3}c_{13}^2- \sqrt{3 s_{13}^2 (2-3 s_{13}^2) -\cos^2 2\theta_\nu}}{\sqrt{3}c_{13}^2+\sqrt{3 s_{13}^2 (2-3 s_{13}^2)-\cos^2 2\theta_\nu}} \hspace{0.2cm}\mbox{for NH},    \\
\frac{\sqrt{3}c_{13}^2+\sqrt{3 s_{13}^2 (2-3 s_{13}^2)-\cos^2 2\theta_\nu}}{\sqrt{3}c_{13}^2-\sqrt{3 s_{13}^2 (2-3 s_{13}^2)-\cos^2 2\theta_\nu}} \hspace{0.2cm}\,\mbox{for  IH}.
\end{array}%
\right. \label{t23sq}\eea

The Jarlskog invariant that relates to the size of CP violation in lepton sector is obtained by comparing the standard parametrization \cite{Pontecorvo1, Pontecorvo2, Jcp, Maki, Rodejohann, Jarlskog1, Jarlskog2, Jarlskog3} and
the model result in Eq. (\ref{Ulep}),
\bea
c_{12} c_{13}^2 c_{23} s_{12} s_{13} s_{23} \sin\delta=\left\{
\begin{array}{l}
 \hspace{0.1 cm}\fr{\cos(2\theta_\nu)}{6\sqrt{3}} \hspace{0.3cm}\mbox{for \, NH},    \\
\hspace{-0.2 cm}-\fr{\cos(2\theta_\nu)}{6\sqrt{3}}  \hspace{0.25cm}\,\mbox{for \, IH}.
\end{array}%
\right. \label{J}
\eea
Equations (\ref{s13sq})-(\ref{J}) yield the following relations:
\bea
&&t^2_{12} =\fr{1}{2 - 3 s_{13}^2} \hspace{0.2cm}\mbox{for \, both NH and IH}, \label{t12s13relation}\\
&& \cos\theta_\nu =\left\{
\begin{array}{l}
\frac{1}{\sqrt{2}}\sqrt{1-\frac{\sqrt{\sin^2 2\theta_{23}-\left(2-3 s_{12}^2\right)^2}}{\sqrt{3} s_{12}^2}} \hspace{0.3cm}\mbox{for  NH},    \\
\frac{1}{\sqrt{2}}\sqrt{1+\frac{\sqrt{\sin^2 2\theta_{23}-\left(2-3 s_{12}^2\right)^2}}{\sqrt{3} s_{12}^2}} \hspace{0.25cm}\,\mbox{for  IH},
\end{array}%
\right.  \label{costhetas12s23}\\
&&\cos\psi=\left\{
\begin{array}{l}
\hspace{0.2 cm}\frac{\sqrt{3} \cos 2\theta_{12}}{\sqrt{4-3\sin^2 2\theta_{12}-\sin^2 2\theta_{23}}} \hspace{0.3cm}\mbox{for  NH},    \\
\hspace{-0.15 cm}-\frac{\sqrt{3} \cos 2\theta_{12}}{\sqrt{4-3\sin^2 2\theta_{12}-\sin^2 2\theta_{23}}} \hspace{0.25cm}\,\mbox{for  IH}.
\end{array}%
\right., \label{cosalphas12s23}\\
&&\sin\de  =\frac{\sqrt{\frac{4}{3}  \cot^2_{12}\left(2t_{12}^2-1\right) \left(t_{23}^4+1\right)-\frac{4}{9}\cot^4_{12}
\left(2 t_{12}^2-1\right)^2 \left(t_{23}^4+t_{23}^2+1\right)-\left(t_{23}^2-1\right)^2}}{\sqrt{3} c_{12}^2 \sin 2 \theta_{23}\sqrt{2 t_{12}^2-1} \left(t_{23}^2+1\right)\left(\frac{2}{3} t_{12}^2 \cot^2_{12}-\frac{1}{3}\cot^2_{12}-1\right)}\crn
&&\hspace{1.3cm}\mbox{for both NH and IH}.  \label{sds12s23}\eea

\section{\label{numerical} Numerical analysis}
\begin{figure}[ht]
\begin{center}
\vspace{-0.85 cm}
\hspace{-6.25 cm}
\includegraphics[width=0.825\textwidth]{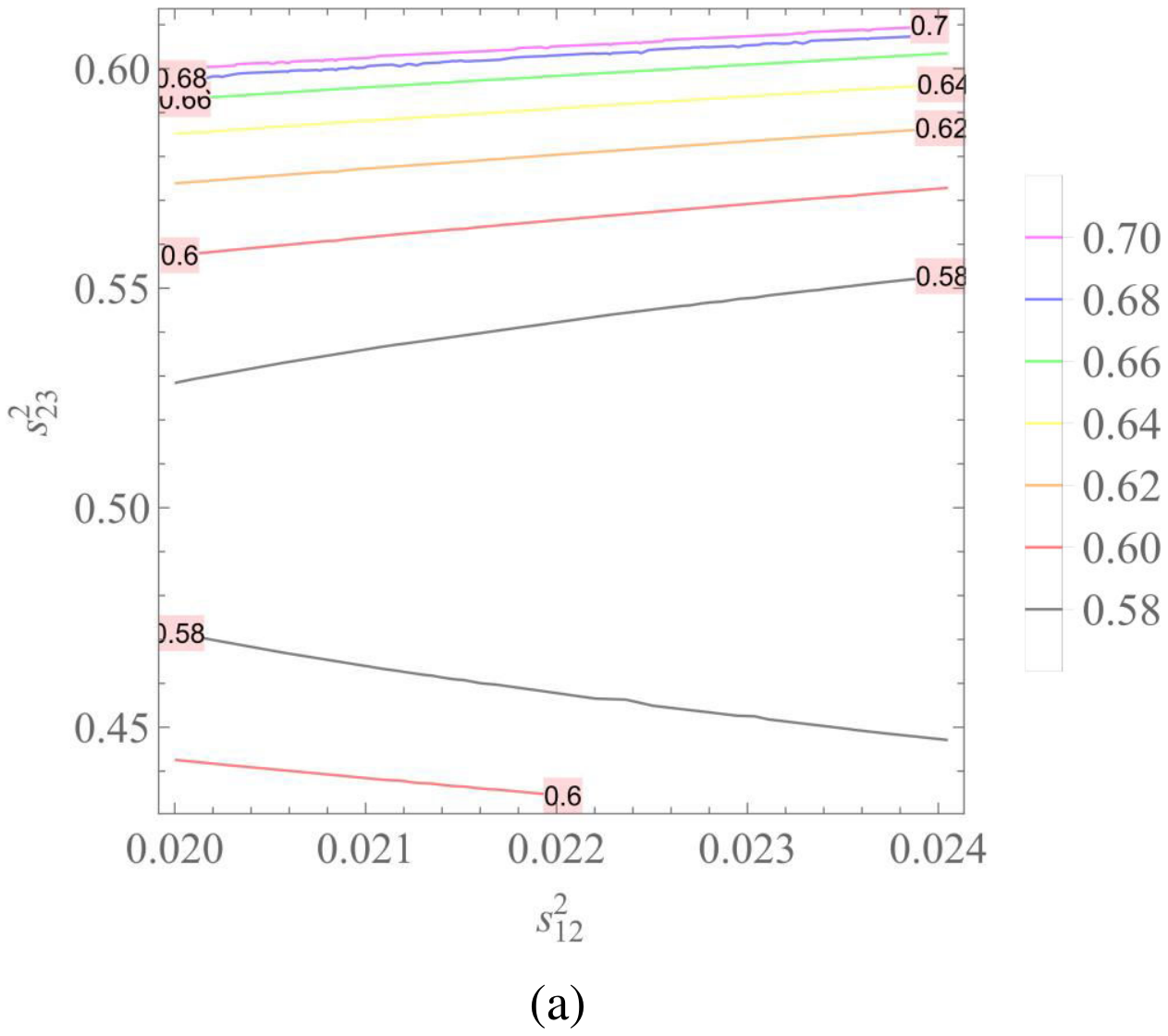}\hspace{-5.25 cm}
\vspace{-0.95 cm}
\includegraphics[width=0.825\textwidth]{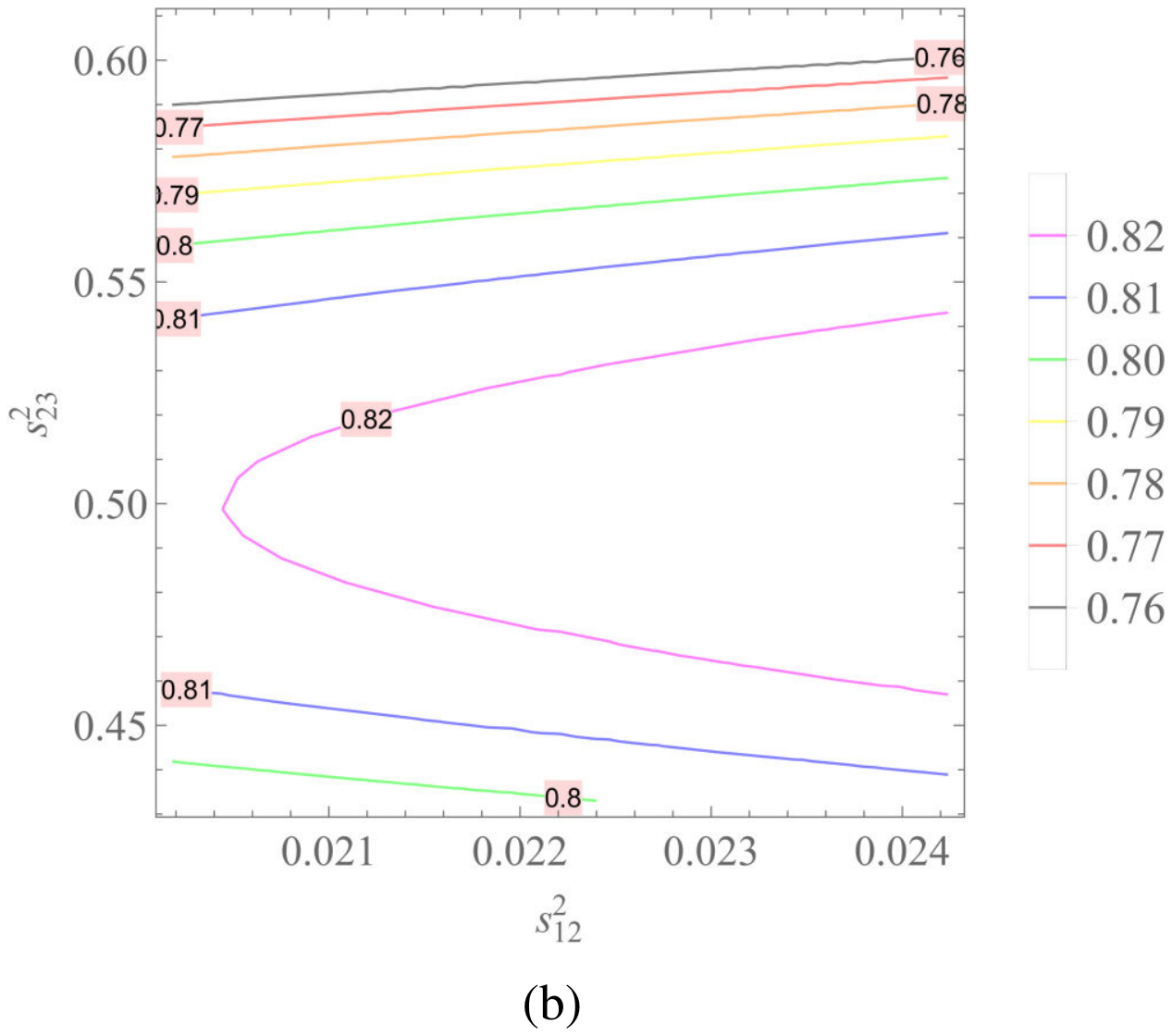}\hspace*{-5.75 cm}
\end{center}
\vspace{-8.5 cm}
\caption[(Colored lines) $\cos\theta_\nu$
versus $s^2_{13}$ and $s^2_{23}$ with (a) $s^2_{13}\in (2.000,\, 2.405) 10^{-2}$ and $s^2_{23}\in (0.434,\, 0.610)$ for NH \cite{Salas2021}, and (b) $s^2_{13}\in (2.018,\, 2.424) 10^{-2}$ and $s^2_{23}\in (0.433,\, 0.608)$ for IH \cite{Salas2021}.]{(Colored lines) $\cos\theta_\nu$
versus $s^2_{13}$ and $s^2_{23}$ with (a) $s^2_{13}\in (2.000,\, 2.405) 10^{-2}$ and $s^2_{23}\in (0.434,\, 0.610)$ for NH \cite{Salas2021}, and (b) $s^2_{13}\in (2.018,\, 2.424) 10^{-2}$ and $s^2_{23}\in (0.433,\, 0.608)$ for IH \cite{Salas2021}.}
\label{costhetaF}
\vspace{-0.25 cm}
\end{figure}

\begin{figure}[ht]
\begin{center}
\vspace{-0.85 cm}
\hspace{-6.25 cm}
\includegraphics[width=0.825\textwidth]{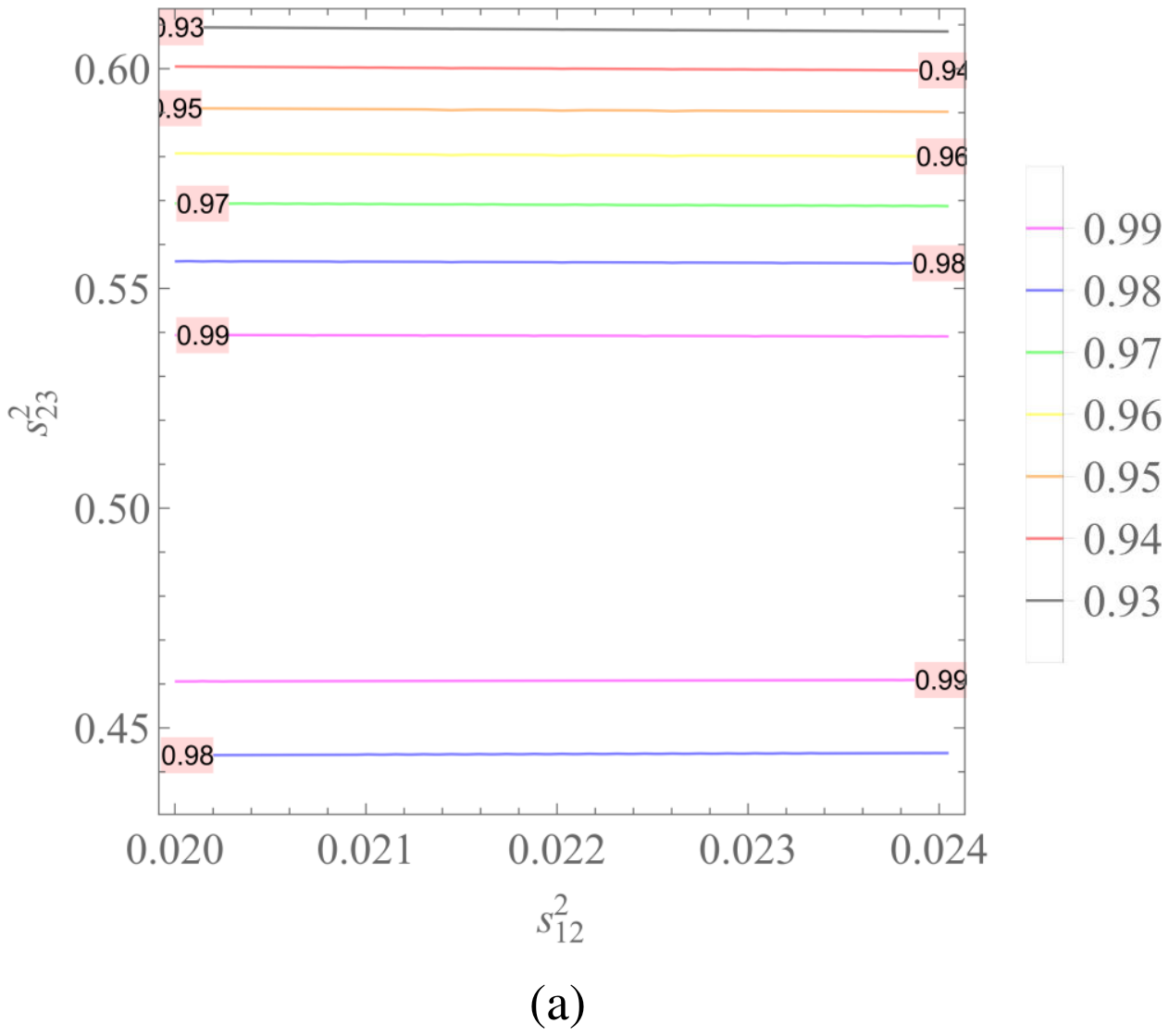}\hspace{-5.25 cm}
\vspace{-0.95 cm}
\includegraphics[width=0.825\textwidth]{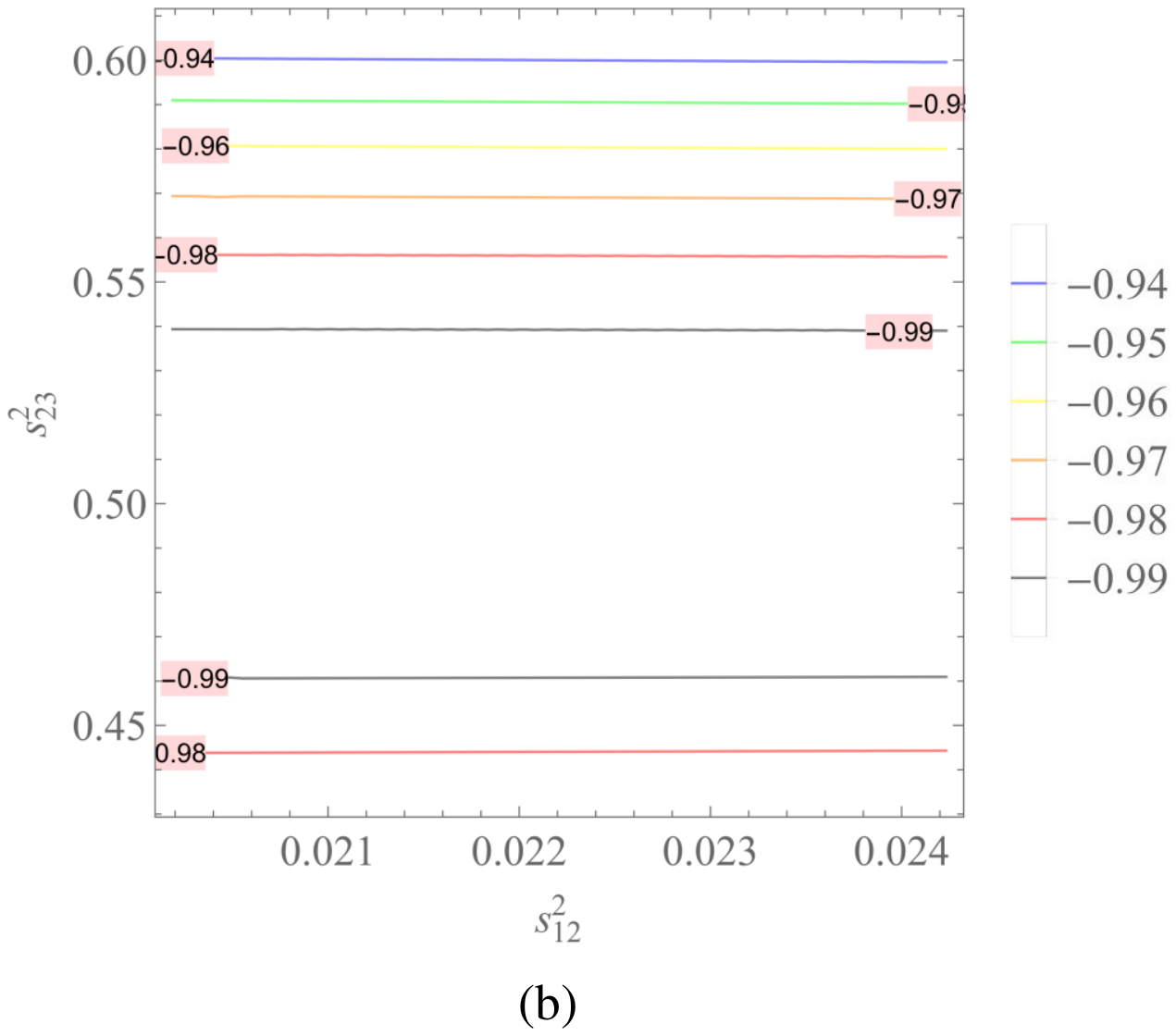}\hspace*{-5.75 cm}
\end{center}
\vspace{-8.5 cm}
\caption[(Colored lines) $\cos\psi$
versus $s^2_{13}$ and $s^2_{23}$ with (a) $s^2_{13}\in (2.000,\, 2.405) 10^{-2}$ and $s^2_{23}\in (0.434,\, 0.610)$ for NH \cite{Salas2021}, and (b) $s^2_{13}\in (2.018,\, 2.424) 10^{-2}$ and $s^2_{23}\in (0.433,\, 0.608)$ for IH \cite{Salas2021}.]
{(Colored lines) $\cos\psi$
versus $s^2_{13}$ and $s^2_{23}$ with (a) $s^2_{13}\in (2.000,\, 2.405) 10^{-2}$ and $s^2_{23}\in (0.434,\, 0.610)$ for NH \cite{Salas2021}, and (b) $s^2_{13}\in (2.018,\, 2.424) 10^{-2}$ and $s^2_{23}\in (0.433,\, 0.608)$ for IH \cite{Salas2021}.}
\label{cospsiF}
\vspace{-0.25 cm}
\end{figure}

\begin{figure}[ht]
\begin{center}
\vspace{-0.85 cm}
\hspace{-6.25 cm}
\includegraphics[width=0.825\textwidth]{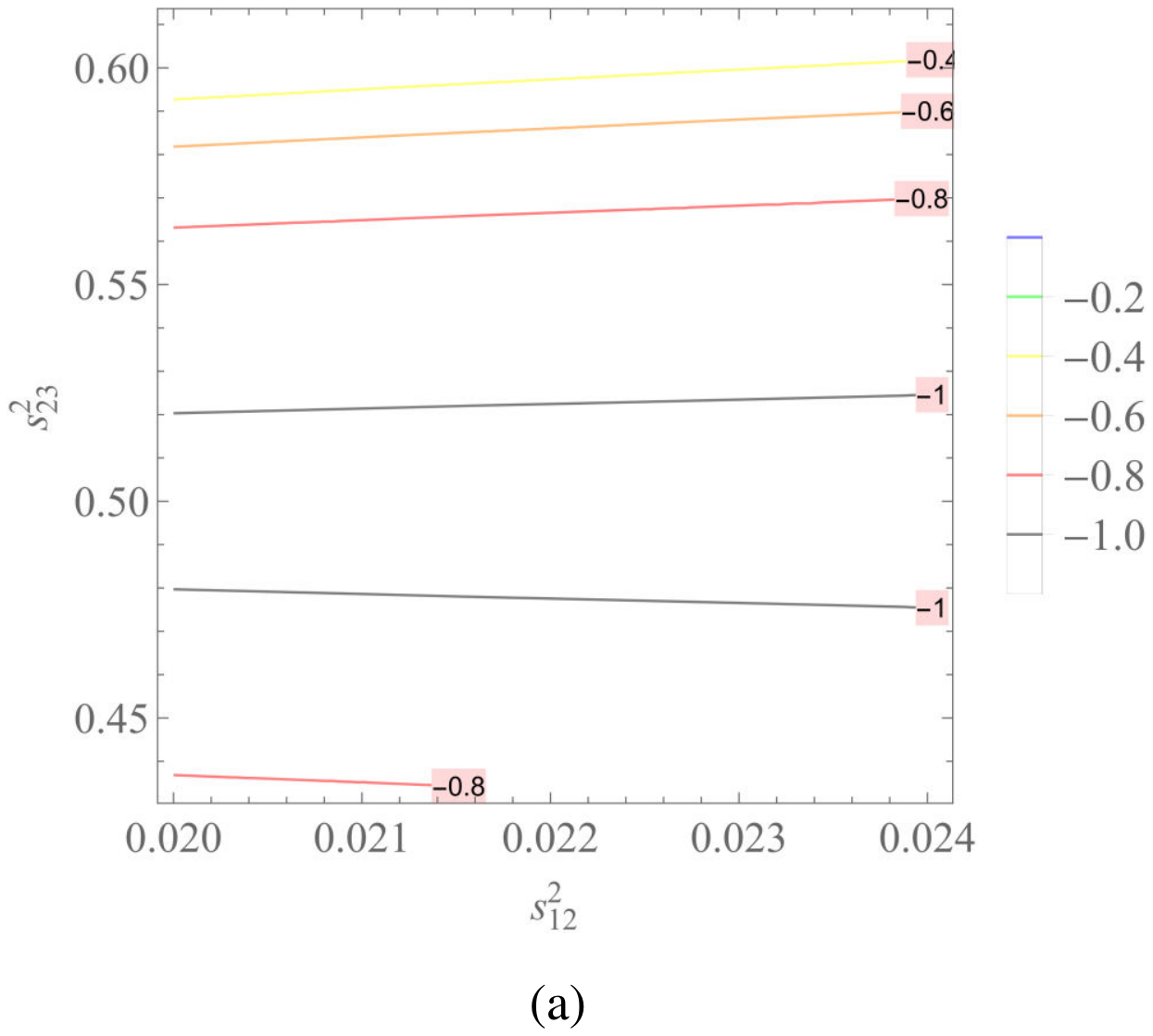}\hspace{-5.5 cm}
\vspace{-0.5 cm}
\includegraphics[width=0.825\textwidth]{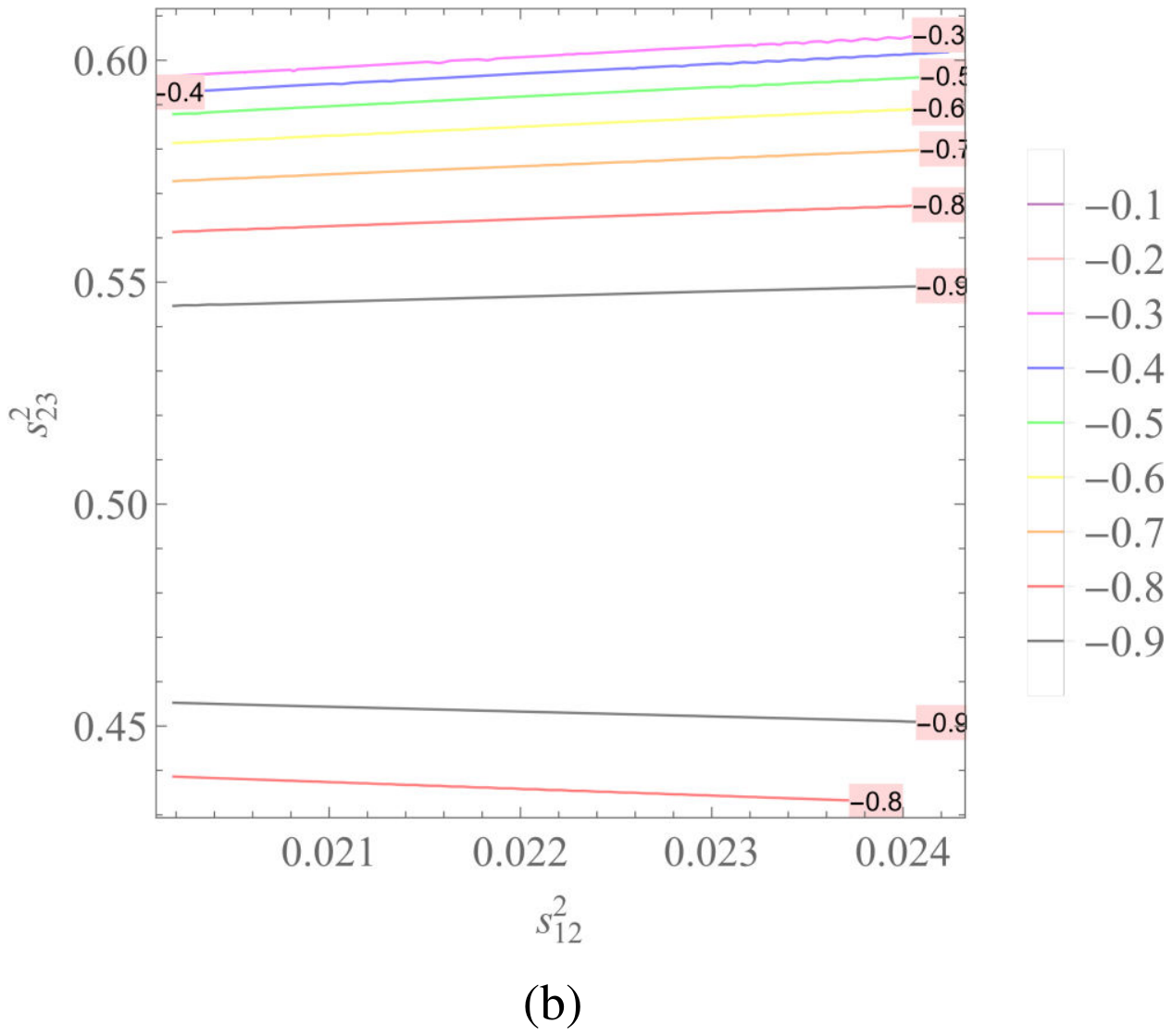}\hspace*{-5.75 cm}
\end{center}
\vspace{-8.85 cm}
\caption[(Colored lines) $\sin\delta$
versus $s^2_{13}$ and $s^2_{23}$ with (a) $s^2_{13}\in (2.000,\, 2.405) 10^{-2}$ and $s^2_{23}\in (0.434,\, 0.610)$ for NH \cite{Salas2021}, and (b) $s^2_{13}\in (2.018,\, 2.424) 10^{-2}$ and $s^2_{23}\in (0.433,\, 0.608)$ for IH \cite{Salas2021}.]
{(Colored lines) $\sin\delta$
versus $s^2_{13}$ and $s^2_{23}$ with (a) $s^2_{13}\in (2.000,\, 2.405) 10^{-2}$ and $s^2_{23}\in (0.434,\, 0.610)$ for NH \cite{Salas2021}, and (b) $s^2_{13}\in (2.018,\, 2.424) 10^{-2}$ and $s^2_{23}\in (0.433,\, 0.608)$ for IH \cite{Salas2021}.}
\label{sindeltaF}
\vspace{-0.5 cm}
\end{figure}

Firstly, the global analysis of the neutrino oscillation data in Ref. \cite{Salas2021} tells us that at $3\, \si $
range of the best-fit value, $s^2_{13}\in (2.00, 2.405)\times10^{-2}$ for NH and $s^2_{13}\in (2.018, 2.424)\times10^{-2}$
for IH. Thus, from  Eq. (\ref{t12s13relation}) we can deduce 
\bea
t^2_{12} \in \left\{
\begin{array}{l}
(0.5155, 0.5187) \hspace{0.2cm}\mbox{for  NH},    \\
(0.5156, 0.5189) \hspace{0.2cm}\mbox{for  IH},
\end{array}%
\right. \,\, \mathrm{i.e.},\,\, \theta_{12}(^\circ) \in \left\{
\begin{array}{l}
(35.68, 35.76) \hspace{0.2cm}\mbox{for  NH},    \\
(35.68, 35.77) \hspace{0.2cm}\mbox{for  IH}.
\end{array}%
\right.\label{t12sqrange}
\eea
Expressions (\ref{costhetas12s23})-(\ref{sds12s23}) show that $\cos\theta_\nu, \cos\psi$ and $\sin\delta$ depend on two parameters $s^2_{13}$ and $s^2_{23}$ which are plotted in Figs. \ref{costhetaF}, \ref{cospsiF} and \ref{sindeltaF}, respectively, within $3\, \si $ range values of $s^2_{13}$ and $s^2_{23}$ \cite{Salas2021}.
\\
Figures \ref{costhetaF} and  \ref{cospsiF} imply that, in $3\, \si $ range, 
 the  ranges of $\cos\theta_\nu$ and  $\cos\psi$ respectively are
\bea
&&\cos\theta_\nu  \in \left\{
\begin{array}{l}
(0.580, 0.700) \hspace{0.3cm}\mbox{for  NH},    \\
(0.760, 0.782) \hspace{0.25cm}\,\mbox{for  IH},
\end{array}%
\right. \,\,\,\,\,\mathrm{i.e.}, \,\, \theta_\nu (^\circ) \in \left\{
\begin{array}{l}
(45.570,\, 54.550) \hspace{0.3cm}\mbox{for  NH},    \\
(38.560,\, 40.540) \hspace{0.25cm}\,\mbox{for  IH}.
\end{array}%
\right. \hs\hs \label{thetanu1}\\
&&\cos\psi \hspace{0.1 cm}\in \left\{
\begin{array}{l}
(0.930, 0.990) \hspace{0.3cm}\mbox{for  NH},    \\
(-0.940, -0.990) \hspace{0.25cm}\,\mbox{for  IH},
\end{array}%
\right. \hspace{0.075 cm}\mathrm{i.e.}, \psi (^\circ) \in \left\{
\begin{array}{l}
(8.110,\, 21.570) \hspace{0.75cm}\mbox{for NH},    \\
(160.100,\, 171.900) \hspace{0.1 cm}\,\mbox{for  IH}.
\end{array}%
\right. \label{psi1}
\eea
Further, Fig.\ref{sindeltaF} implies that, in $3\, \si $ range, the Dirac CP phase is predicted to be
\bea
\sin\de  \in \left\{
\begin{array}{l}
(-1.00, -0.20) \hspace{0.25cm}\mbox{for  NH},    \\
(-0.90, -0.10) \hspace{0.25cm}\mbox{for  IH},
\end{array}%
\right. \,\, \mathrm{i.e.}, \,\,  \delta (^\circ)  \in \left\{
\begin{array}{l}
(270.00,\, 348.50) \hspace{0.3cm}\mbox{for  NH},    \\
(295.80,\, 354.30) \hspace{0.25cm}\,\mbox{for  IH},
\end{array}%
\right. \label{delta1}
\eea
which all belong to $3\, \si $ range of the best-fit value taken from Ref. \cite{Salas2021}.\\
The ranges on the magnitude of the elements of the leptonic mixing \ref{Ulep} are obtained as
\bea
|U_{\mathrm{lep}}|\in \left\{
\begin{array}{l}
\left(%
\begin{array}{ccc}
0.800-0.804    &\hs 0.577 \hs&0.133-0.155  \\
0.440-0.580  &\hs 0.577 \hs&0.620-0.690 \\
0.225-0.400&\hs 0.577 \hs& 0.700-0.770 \\
\end{array}%
\right) \hspace{0.2cm}\mbox{for  NH},    \\
\left(%
\begin{array}{ccc}
0.802-0.806    &\hs 0.577 \hs&0.145-0.165  \\
0.280-0.420  &\hs 0.577 \hs&0.700-0.780 \\
0.440-0.520&\hs 0.577 \hs& 0.600-0.680 \\
\end{array}%
\hspace{-0.05cm}\right)\hspace{0.25cm}\mbox{for  IH.}
\end{array}%
\right. \label{Ulepranges}
\vspace{-0.5 cm}
\eea
In the case of $s^2_{13}$ and $s^2_{23}$ get their best-fi values taken from Ref. \cite{Salas2021} as shown in Table \ref{Salas2021}, i.e., $s^2_{13}=2.20\times 10^{-2}\, (\theta_{13} =8.53^\circ),\, s^2_{23}=0.574 \, (\theta_{23} =49.26^\circ)$ for NH while $s^2_{13}=2.225\times 10^{-2}\, (\theta_{13} =8.58^\circ),\, s^2_{23}=0.578 \, (\theta_{23} =49.49^\circ)$ for IH, we get $s^2_{12} = 0.341\, (\theta_{12} =35.72^\circ)$, and
\bea
&&\cos\theta_\nu = \left\{
\begin{array}{l}
0.611 \hspace{0.3cm}\mbox{for \, NH},    \\
0.788 \hspace{0.25cm}\,\mbox{for \, IH},
\end{array}%
\right. \,\,\, \mathrm{i.e.,} \,\,\,\, \theta_\nu (^\circ) =\left\{
\begin{array}{l}
52.370 \hspace{0.3cm}\mbox{for  NH},    \\
37.960 \hspace{0.25cm}\,\mbox{for  IH}.
\end{array}%
\right. \label{thetanubf}\\
&&\cos\psi \hspace{0.05 cm}=\hspace{0.05 cm} \left\{
\begin{array}{l}
0.966 \hspace{0.3cm}\mbox{for \, NH},    \\
\hspace{-0.15cm}-0.962 \hspace{0.1cm}\,\mbox{for \, IH},
\end{array}%
\right. \,\,\, \mathrm{i.e.,} \,\,\,\, \psi (^\circ) \hspace{0.035 cm}=\hspace{0.035 cm}\left\{
\begin{array}{l}
15.030 \hspace{0.3cm}\mbox{for  NH},    \\
164.20 \hspace{0.2cm}\,\mbox{for  IH},
\end{array}%
\right. \label{psibf}\\
&&\sin\de \hspace{0.1 cm}=\hspace{0.1 cm} \left\{
\begin{array}{l}
\hspace{-0.15cm}-0.737 \hspace{0.2cm}\mbox{for \, NH},    \\
\hspace{-0.15cm}-0.686 \hspace{0.15cm}\,\mbox{for \, IH},
\end{array}%
\right. \,\,\mathrm{i.e.,} \,\,\,\,  \delta (^\circ)\hspace{0.075cm}=\hspace{0.075cm}\left\{
\begin{array}{l}
312.60 \hspace{0.25cm}\mbox{for  NH},    \\
316.7 \hspace{0.35cm}\,\mbox{for IH}.
\end{array}%
\right. \label{deltabf}
\eea
The obtained values of $\delta$ in Eq. (\ref{deltabf}) belong to $3\, \si $ range of the best-fit value \cite{Salas2021}.\\
As a {result, the Jarlskog invariant is obtained as follows
\bea
J&=& \left\{
\begin{array}{l}
-2.449\times 10^{-2} \hspace{0.3cm}\mbox{for \, NH},    \\
-2.342\times 10^{-2}\hspace{0.25cm}\,\mbox{for \, IH}.
\end{array}%
\right. \hspace{0.675cm}
\eea
By above analysis we can conclude that the considered model can explain the observed pattern of lepton mixing \cite{Salas2021}
 in which the reactor and atmospheric angles
get the best-fit values, and
the solar angle and Dirac phase lie within $3\,\si $ limits for both NH and IH.

We now turn to neutrino mass hierarchy. Comparing neutrino mass obtained from the model in Eqs. (\ref{m123sq}) and (\ref{Unu}) with the best-fit values of the neutrino mass-squared differences taken from Ref. \cite{Salas2021} as shown in Tab. \ref{Salas2021},  $\De  m^2_{21}=75.0\, \mathrm{meV}^2$ and $\De  m^2_{31}=2.55\times 10^{3}\, \mathrm{meV}^2$ for NH while $\De  m^2_{31}=-2.45\times 10^{3}\, \mathrm{meV}^2$ for IH,  we get a solution
\bea
&&\Gamma_1 =m^2_{2}-\Delta m^2_{21}+\frac{\Delta m^2_{31}}{2} \hspace{0.2 cm}\mbox{(both  NH and IH)}, \hs \Gamma_2 = \left\{
\begin{array}{l}
\hspace{0.1cm}\frac{\Delta m^2_{31}}{2}  \hspace{0.3cm}\mbox{for  NH},    \\
\hspace{-0.15cm}-\frac{\Delta m^2_{31}}{2}\hspace{0.2cm}\mbox{for  IH},
\end{array}%
\right. \label{Ga12}\\
&&m_1=\sqrt{m^2_{2}-\Delta m^2_{21}}, \,\, m_3=\sqrt{m^2_{2}-\Delta m^2_{21}+\Delta m^2_{31}} \hs \mbox{(both  NH and IH)}, \label{m1m3} \\
&&\sum m_\nu=m_{2} + \sqrt{m^2_{2}-\Delta m^2_{21}} + \sqrt{m^2_{2}-\Delta m^2_{21} +\Delta m^2_{31}} \hs \mbox{(both  NH and IH)}. 
\label{sum}
\eea
Expressions (\ref{Ga12})-(\ref{sum}) show that $m_1$ depends on two parameters $m^2_2\equiv A^2_{02}$ and $\Delta m^2_{21}$ while $m_3$ and $\sum m_\nu$ depend on three parameters $m^2_2$, $\Delta m^2_{21}$ and $\Delta m^2_{31}$. At present the absolute value of the neutrino mass remains unknown, however, the KATRIN Collaboration has reported an upper limit on the neutrino mass of $m_\nu < 1.1\, \mathrm{eV}$ \cite{Katrin21PRL, Katrin21PRD} or an improved upper limit of $m_\nu < 0.8\, \mathrm{eV}$ \cite{Katrin21arX}. Thus, in order to determine the neutrino mass hierarchy, the sum of neutrino mass as well as the effective neutrino masses in Section \ref{effectivemass}, we will consider $A^2_{02}\equiv m^2_2, \Delta m^2_{21}$ and $\Delta m^2_{31}$ as input parameters with\footnote{For NH $m^{2}_2$ must be larger than $75\, \mathrm{meV}^{2}$ because the real condition of $m_1$, and for IH $m^{2}_2$ must be larger than $2.525\times 10^3\, \mathrm{meV}^{2}$ because the real condition of $m_3$. In this work, we consider $m^2_{2} \in (10^2, 10^{3}) \, \mathrm{meV}^2$ for NH and $m^2_{2} \in (2.55\times 10^3, 10^4) \, \mathrm{meV}^2$ for IH because the normal mass spectrum is achieved with $m^2_{2} \in (10^2, 10^{3}) \, \mathrm{meV}^2$ and the inverted mass spectrum is achieved with $m^2_{2} \in (2.55\times 10^3, 10^4) \, \mathrm{meV}^2$. In the case of $m^2_{2} >10^{3}$ for NH and $m^2_{2}> 10^4 \, \mathrm{meV}^2$ for IH, the neutrino spectrum will be nearly degenerate.} $m^2_2 \in (10^2, 10^{3})\, \mathrm{meV}^2$ for NH and $m^2_2 \in (2.55\times 10^3, 10^{4})\, \mathrm{meV}^2$ for IH while $\Delta m^2_{21}$ and $\Delta m^2_{31}$ get their best-fit values taken from \cite{Salas2021}, i.e.,  $\Delta m^2_{21}=75 \,\mathrm{meV}^2$ and $\Delta m^2_{31} =2.55\times 10^3 \,\mathrm{meV}^2$ for NH while $\Delta m^2_{31}=-2.45\times10^3 \,\mathrm{meV}^2$ for IH.
At the best-fit points of $\Delta m^2_{21}$ and $\Delta m^2_{31}$, the dependence of $m_{1,3}$ on $m^2_2$  is depicted in Fig. \ref{m1m3F} which implies
\bea
\left\{
\begin{array}{l}
m_1\in (5.00, 30.40)\, \mathrm{meV},\hs m_3\in (50.74, 58.95)\, \mathrm{meV} \hspace{0.15cm}\mbox{for  NH},    \\
m_1\in (49.75, 99.62)\, \mathrm{meV},\, m_3\in (5.00, 86.46)\, \mathrm{meV} \hspace{0.275cm}\mbox{for  IH}.
\end{array}%
\right. \label{sumrange}
\eea
\begin{figure}[ht]
\begin{center}
\vspace*{-0.5cm}
\hspace*{-1.75cm}
\includegraphics[width=0.675\textwidth]{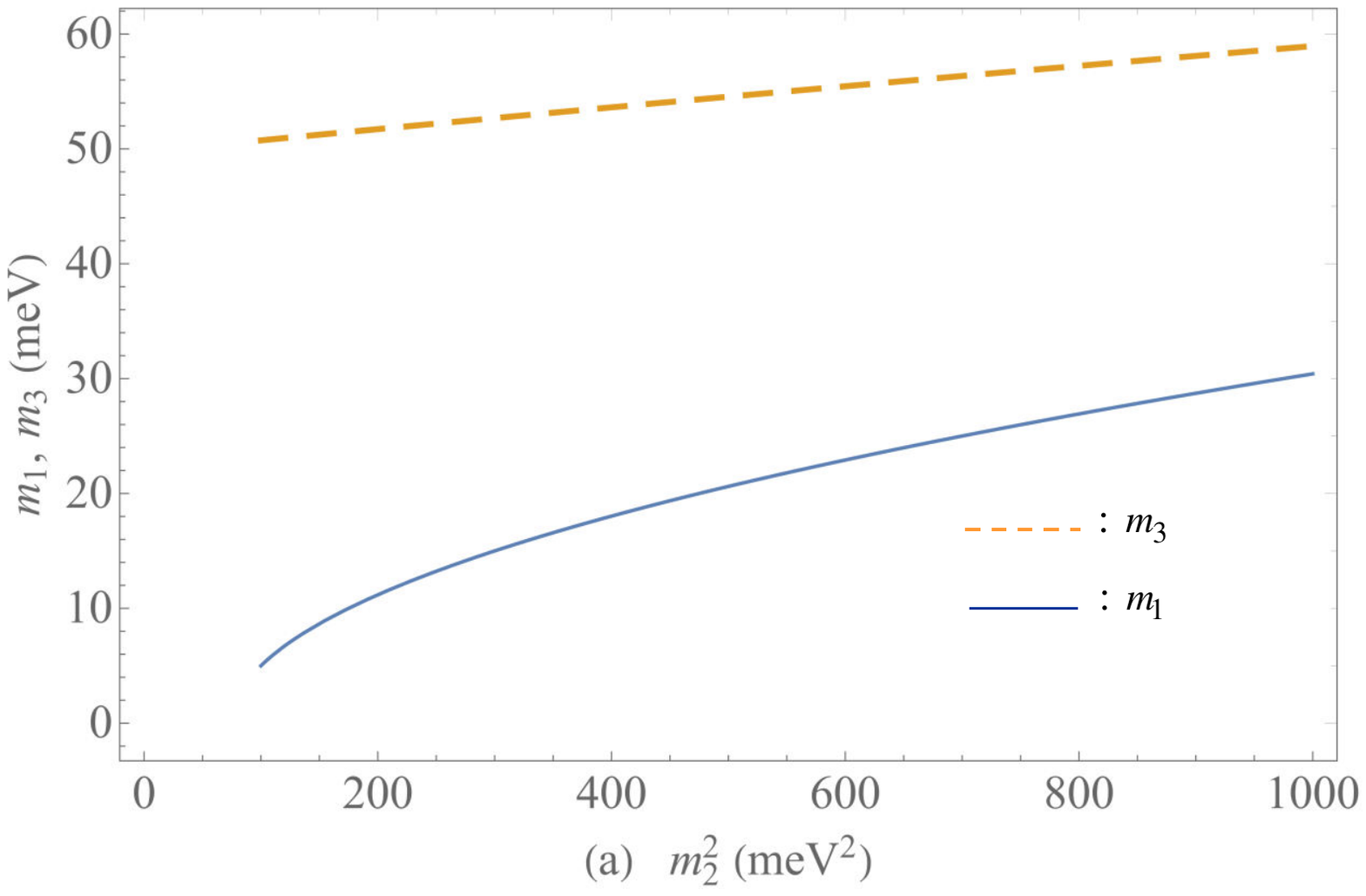}\hspace*{-2.75cm}
\includegraphics[width=0.675\textwidth]{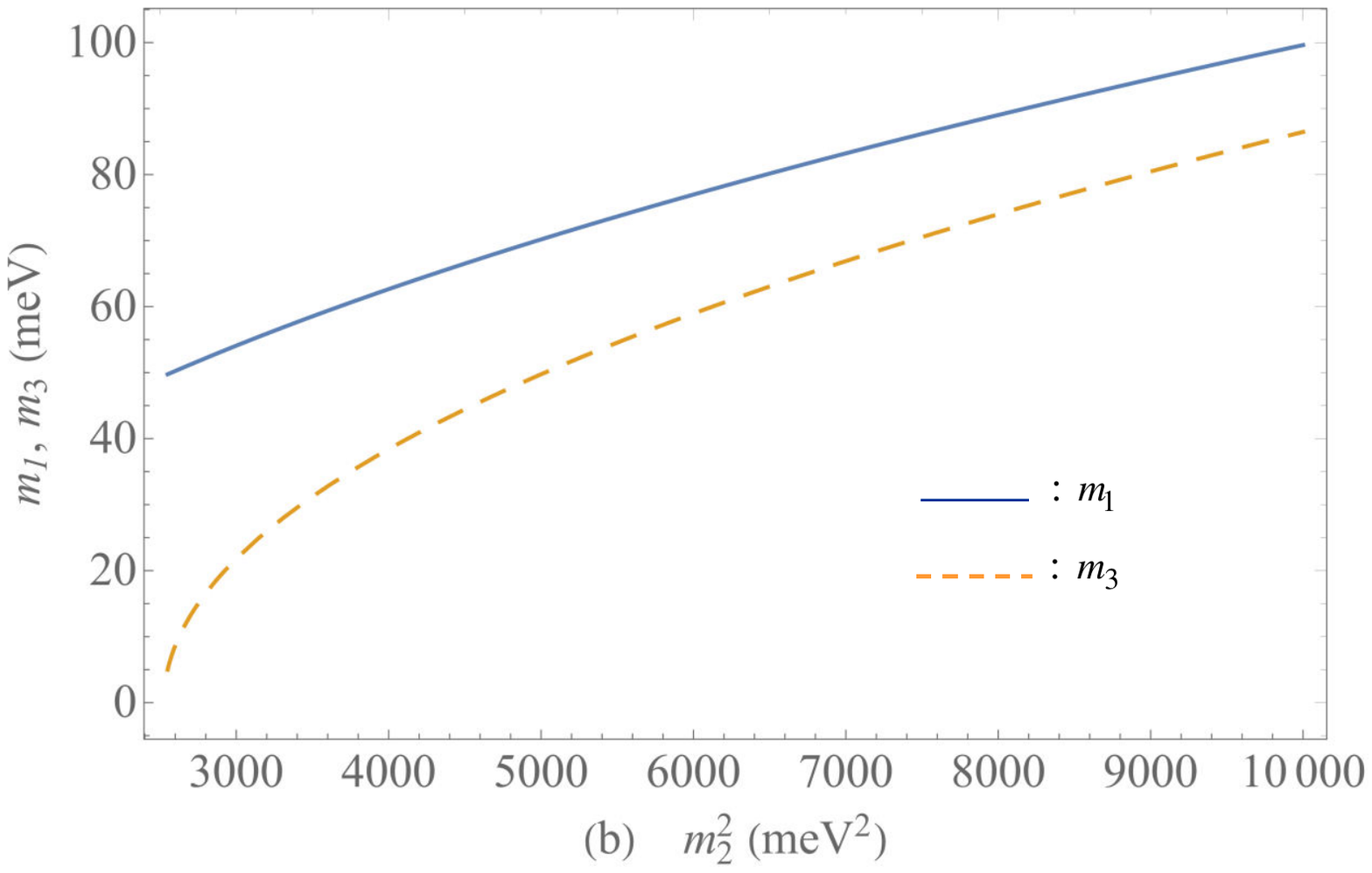} \hspace*{-2.5cm}
\vspace*{-7.85cm}
\caption[$m_1$ and $m_3$\, (in meV) versus $A^2_{02}\equiv m^2_{2}$ with (a) $m^2_{2} \in (10^2, 10^{3}) \,\mathrm{meV}^2$ for NH, and (b) $m^2_{2} \in (2.55\times 10^3, 10^{4}) \, \mathrm{meV}^2$ for IH.]{$m_1$ and $m_3$\, (in meV) versus $A^2_{02}\equiv m^2_{2}$ with (a) $m^2_{2} \in (10^2, 10^{3}) \,\mathrm{meV}^2$ for NH, and (b) $m^2_{2} \in (2.55\times 10^3, 10^{4}) \, \mathrm{meV}^2$ for IH.}\label{m1m3F}
\end{center}
\end{figure}

The sum of neutrino mass as a function of $m^2_2$ is depicted in Fig. \ref{SumF} which implies
\bea
\sum_\nu  m_\nu&\in & \left\{
\begin{array}{l}
(65.74, 121.00)\, \mathrm{meV} \hspace{0.3cm}\mbox{for  NH},    \\
(105.20, 286.10)\, \mathrm{meV}\hspace{0.15cm}\mbox{for  IH}.
\end{array}%
\right. \label{sumrange}
\eea
\begin{figure}[h]
\begin{center}
\hspace*{-0.25cm}
\includegraphics[width=0.5\textwidth]{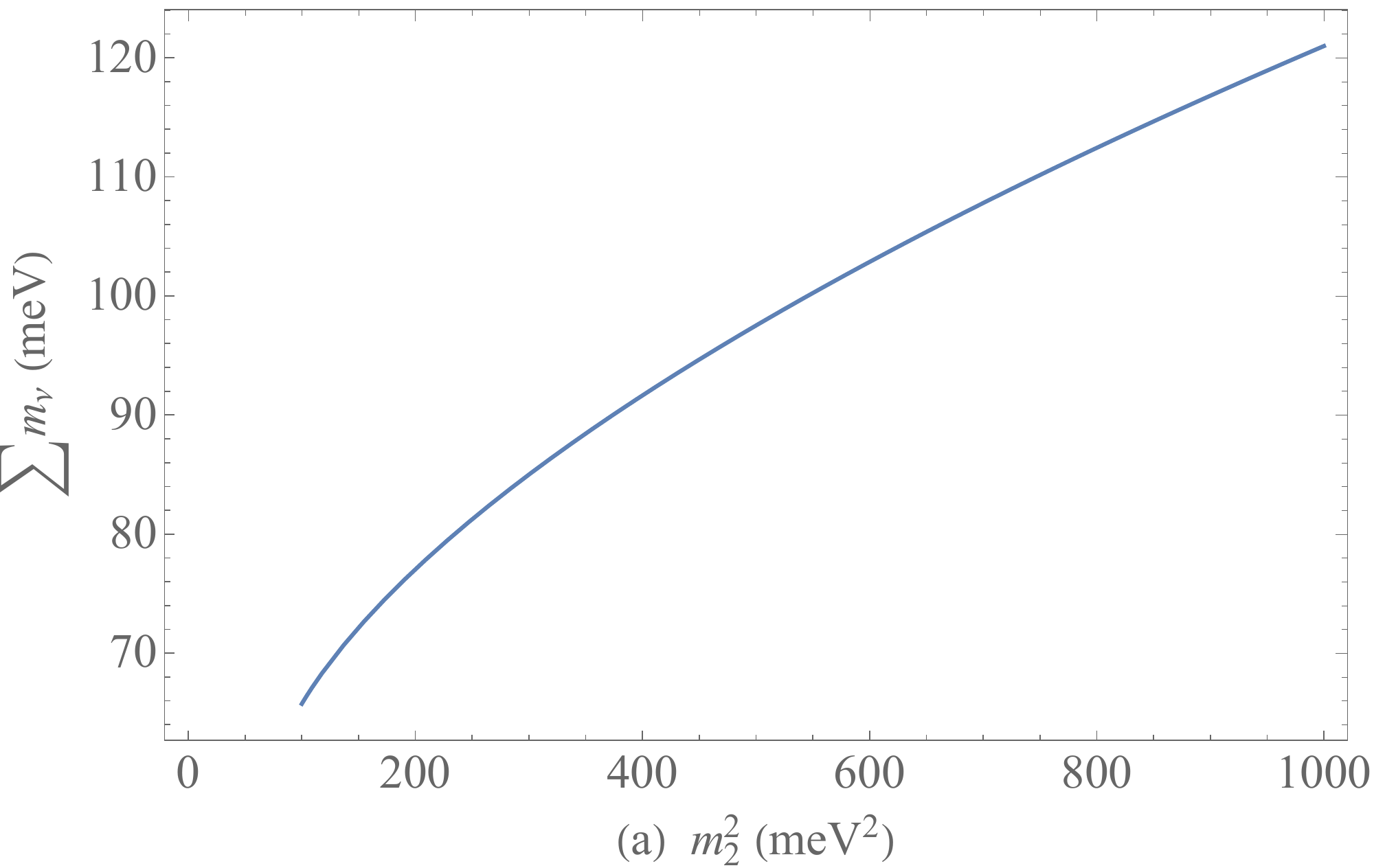}\hspace*{0.12cm}
\includegraphics[width=0.5\textwidth]{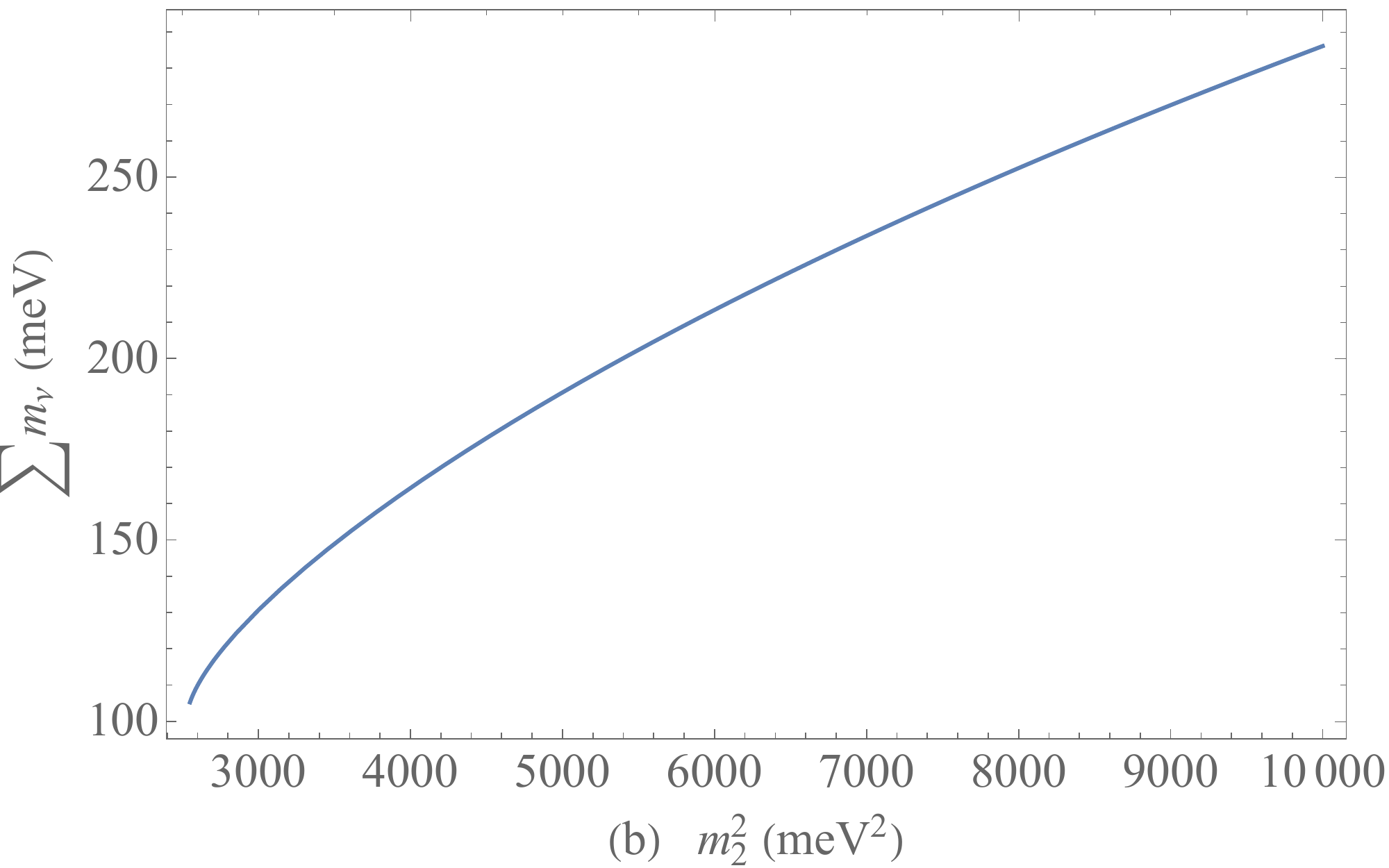}
\vspace*{-0.5 cm}
\caption[The sum of neutrino mass $\sum_\nu  m_\nu$ (in meV) versus $A^2_{02}\equiv m^2_{2}$ with (a) $m^2_{2} \in (10^2, 10^{3}) \,\mathrm{meV}^2$ for NH, and (b) $m^2_{2} \in (2.55\times 10^3, 10^4) \, \mathrm{meV}^2$ for IH.]{The sum of neutrino mass $\sum_\nu  m_\nu$ (in meV) versus $A^2_{02}\equiv m^2_{2}$ with (a) $m^2_{2} \in (10^2, 10^{3}) \,\mathrm{meV}^2$ for NH, and (b) $m^2_{2} \in (2.55\times 10^3, 10^4) \, \mathrm{meV}^2$ for IH.}\label{SumF}
\end{center}
\end{figure}
In order to get the explicit values of the model parameters, we fix $m^2_{2}=10^2\, \mathrm{meV}^2$ for NH and $m^2_{2}=2.55\times 10^3\, \mathrm{meV}^2$ for IH, we then obtain
\bea
&& \left\{
\begin{array}{l}
m_1= 5.00\, \mathrm{meV}, \hspace{0.2 cm} m_2= 10.00\, \mathrm{meV}, \hspace{0.2 cm} m_3=50.74\,\mathrm{meV} \hspace{0.175cm}\mbox{for  NH},    \\
m_1= 49.75\, \mathrm{meV}, \hspace{0.1 cm} m_2= 50.50\, \mathrm{meV}, \hspace{0.2 cm} m_3=5.00\,\mathrm{meV}\hspace{0.35cm}\mbox{for  IH}, \label{m1m2m3bf}
\end{array}
\right.\eea
and
\bea
&&\sum_\nu  m_\nu =\left\{
\begin{array}{l}
65.74 \, \mbox{meV} \hspace{0.35cm} \mbox{for  NH,} \\
105.20\, \mbox{meV} \hspace{0.15cm} \mbox{for  IH}.
\end{array}%
\right.   \label{sumbf}
\eea
There are currently various limits on the sum of neutrino
mass, such as
 $\sum_\nu m_{\nu} < 0.13\,  \mathrm{eV}$ for NH, $\sum_\nu m_{\nu} < 0.15\,  \mathrm{eV}$ for IH \cite{Salas2021} and $\sum_\nu m_\nu < 0.12\div 0.69 \, \mathrm{eV}$ \cite{Capozzi20, nuboundAghanim20, nuboundShadab21}. Thus, the sum of
 neutrino mass predicted by our model in Eq. (\ref{sumrange}) is in well agreement with the most recent experimental limits.
\section{\label{effectivemass}Effective neutrino mass parameter}
Now, we deal with the effective neutrino mass governing the beta decay given by\cite{betdecay2, betdecay4,betdecay5}
\bea
&&m_{\beta } =\sqrt{\left|U_{11 }\right|^2 m_1^2+\left|U_{12}\right|^2 m_2^2+\left|U_{13}\right|^2 m_3^2}. \label{meff}\eea
Here $m_{1,2,3}$ correspond to the masses of three light neutrinos defined in Eqs. (\ref{m123sq}) and (\ref{Unu}) while $U_{11, 12, 13}$ are the leptonic mixing matrix elements given in Eq. (\ref{Ulep}). Expressions (\ref{m123sq}), (\ref{Unu})-(\ref{Ulep}), (\ref{costhetas12s23})-(\ref{cosalphas12s23}), (\ref{m1m3}) and (\ref{meff}) show that the effective neutrino mass $m_{\beta }$ depends on five parameters including $\theta_{12}, \theta_{23}$, $m^2_2$, $\Delta m^2_{21}$ and $\Delta m^2_{31}$.
For NH $m_{light}\equiv m_1$ is the lightest neutrino mass while for IH $m_{light}\equiv m_3$ is the lightest neutrino mass. At the best-fit points of $s^2_{13}$, $s^2_{23}$, $\Delta m^2_{21}$ and $\Delta m^2_{31}$ taken from Ref. \cite{Salas2021} as shown in Table \ref{Salas2021}, i.e., $\Delta m^2_{21}=75 \,\mathrm{meV}^2$ and $\Delta m^2_{31} =2.55\times 10^3 \,\mathrm{meV}^2,\, s^2_{13} = 2.20\times 10^{-2},\, s^2_{23}=0.574$ for NH while $\Delta m^2_{31}=-2.45\times 10^3 \,\mathrm{meV}^2, \,s^2_{13} = 2.225\times 10^{-2},\, s^2_{23}=0.578$ for IH, the effective mass $m_{\beta }$ depends on $m^2_2$ which is plotted in Fig. \ref{mbF}.
\begin{figure}[h]
\begin{center}
\vspace*{-1.25 cm}
\hspace*{-1.75cm}
\includegraphics[width=0.675\textwidth]{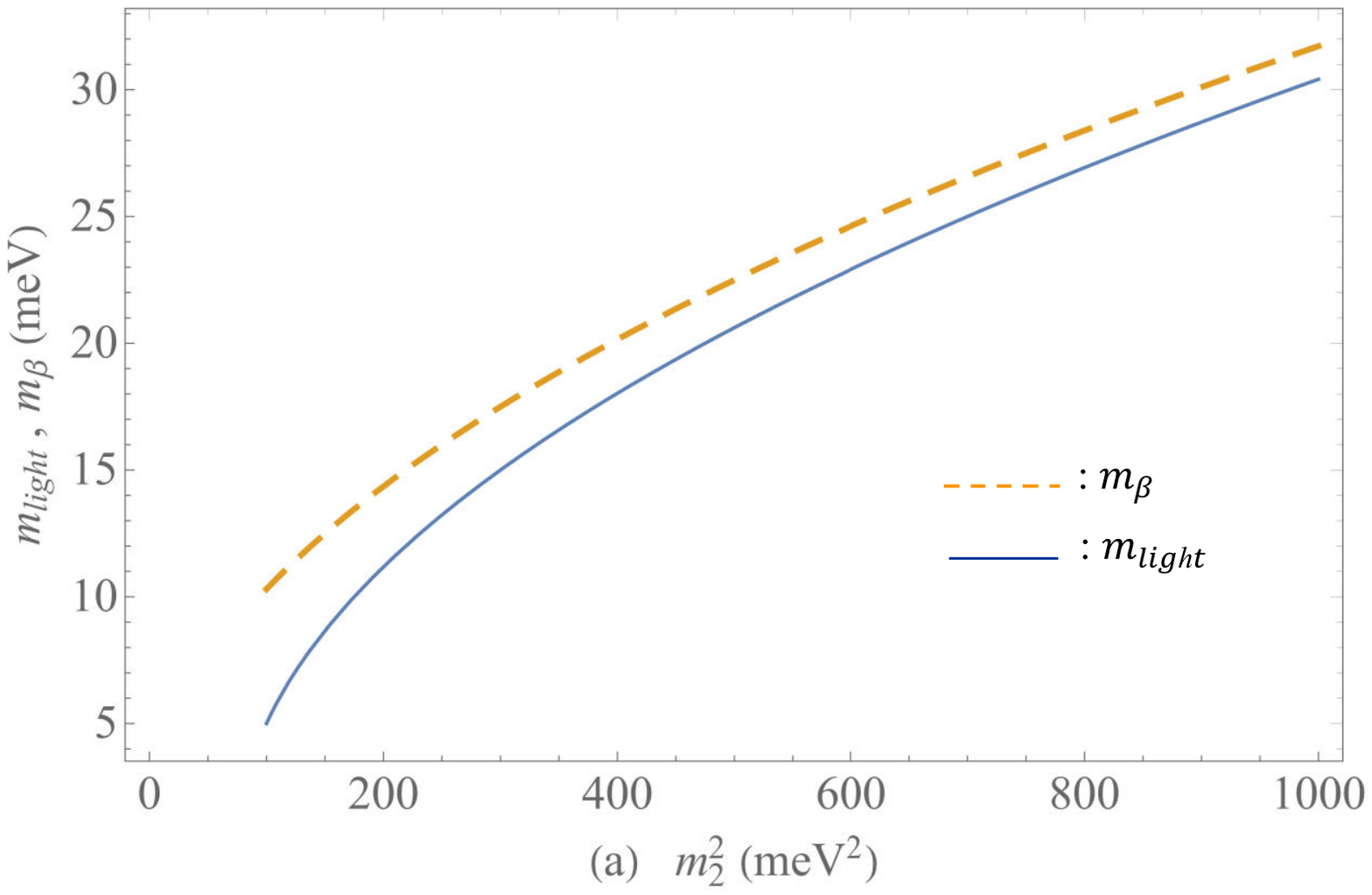}\hspace*{-2.75cm}
\includegraphics[width=0.675\textwidth]{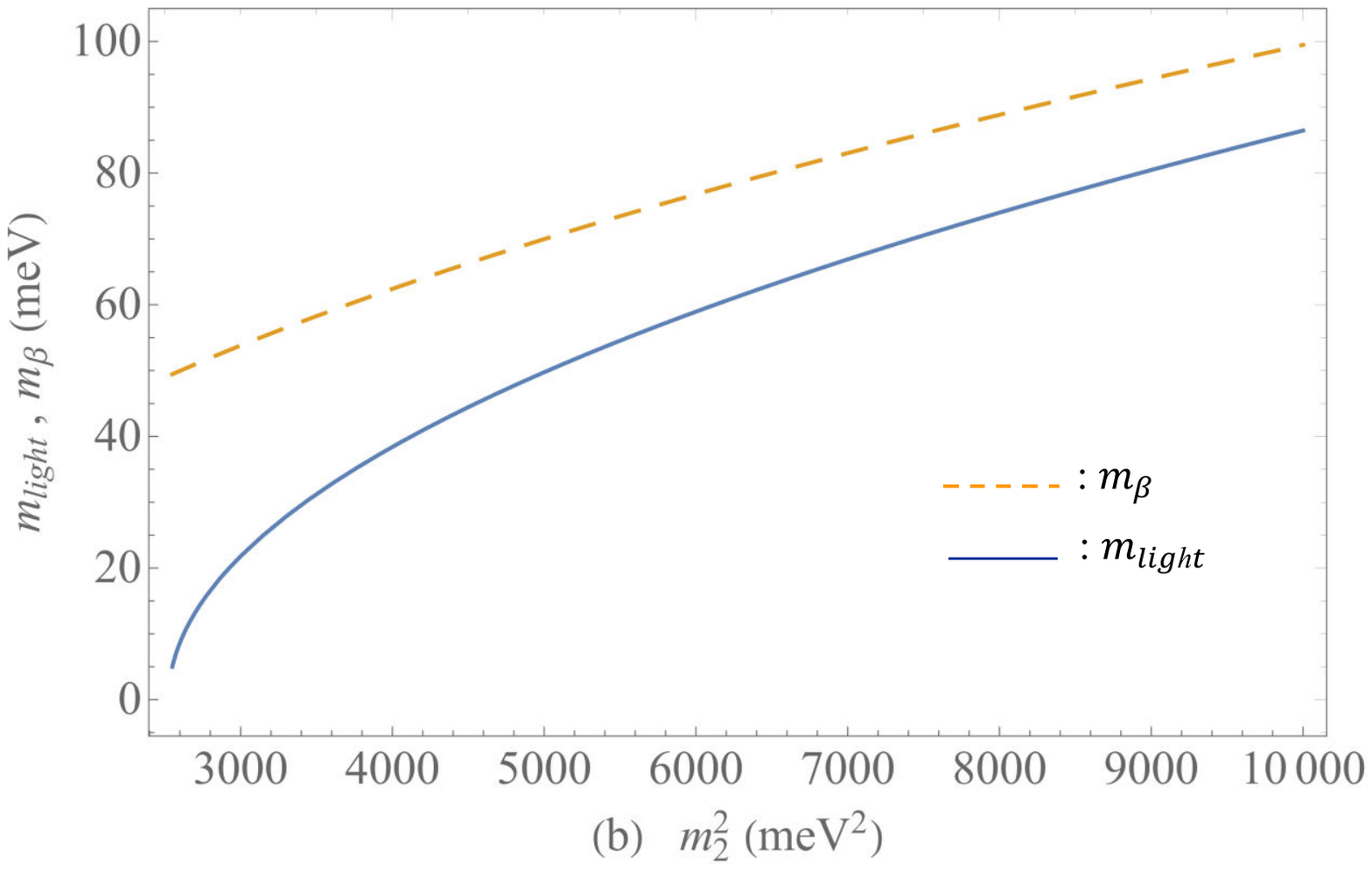}
\vspace*{-8.0cm}
\caption[$m_{light}$ and $m_\beta$\, (in meV) versus $m^2_{2}$ with (a) $m^2_{2} \in (10^2, 10^3) \,\mathrm{meV}^2$ for NH, and (b) $m^2_{2} \in (2.55\times 10^3, 10^4) \, \mathrm{meV}^2$ for IH.]{$m_{light}$ and $m_\beta$\, (in meV) versus $m^2_{2}$ with (a) $m^2_{2} \in (10^2, 10^3) \,\mathrm{meV}^2$ for NH, and (b) $m^2_{2} \in (2.55\times 10^3, 10^4) \, \mathrm{meV}^2$ for IH.}\label{mbF}
\vspace*{-0.5cm}
\end{center}
\end{figure}
This figure implies
\bea
&&m_{\beta} \in\left\{
\begin{array}{l}
(10.30, 31.72)\, \mbox{meV} \hspace{0.25cm} \mbox{for  NH,} \\
(49.45, 99.48)\, \mbox{meV} \hspace{0.25cm} \mbox{for  IH}.
\end{array}%
\right. \label{mbrange}
\eea
Using the model parameters obtained in Section \ref{numerical}, we get the following values:
\bea
&& m_{\beta} =\left\{
\begin{array}{l}
10.30\, \mbox{meV} \hspace{0.25cm} \mbox{for  NH,} \\
49.45\, \mbox{meV} \hspace{0.25cm} \mbox{for  IH}.
\end{array}%
\right.   \label{mbbf}
\eea
The predicted values of the effective neutrino mass in Eq. (\ref{mbrange}), for both NH and IH, satisfy all the upper limits arising from the $\beta$ decay \cite{PDG2020} with $8.5 \,\mathrm{meV} < m_{\beta} < 1.1\, \mathrm{eV}$ for NH and $48 \, \mathrm{meV} < m_{\beta} < 1.1\, \mathrm{eV}$ for IH,  $m_{\beta} < 40\, \mathrm{meV}$ \cite{EsfahaniP8,Esfahani21}, and $m_{\beta} < 0.8\, \mathrm{eV}$ \cite{Aker22n}.
\section{\label{conclusion}Conclusions}
We have constructed a low-scale model with $A_4\times Z_4 \times Z_2$ symmetry and a global lepton
 number $U(1)_L$ symmetry capable of generating the recent neutrino oscillation data.
The model contains only one scalar  doublet and eight scalar singlets. The small neutrino masses are reproduced by linear seesaw
 mechanism. The model can explain the observed pattern of lepton mixing 
in which the reactor and atmospheric angles
get the best-fit values, and
the solar angle and Dirac phase lie within $3\,\si $ limits. The model predicts the Dirac CP phase to lie in $270.00 \leq \delta (^\circ)\leq  348.50$ for NH and  $295.80 \leq \delta^{(\circ)}\leq 354.30$ for IH that includes the currently preferred maximal value.
The obtained values of the sum of neutrino mass and the effective neutrino mass are respectively predicted to be $\sum_\nu  m_\nu \in (65.74, 121.00)\, \mathrm{meV}$ and $m_{\beta}\in (10.30, 31.72) \, \mbox{meV}$ for NH while $\sum_\nu  m_\nu \in (105.20, 286.10)\, \mathrm{meV}$ and $m_{\beta} \in (49.45, 99.48) \, \mbox{meV}$ for IH, which are all below the present experimental limits.
\section*{Acknowledgments}
This work is funded by the
Vietnam Academy of Science and Technology under grant  NVCC05.17/22-22.
\appendix
\section{\label{Appenalbeta} The explicit expressions of $\al_{1,2,3}$ and $\beta_{1,2,3}$ and $\gamma_0$ }
\vspace*{-0.5 cm}
\bea
\al_1&=&\frac{\al_{0}}{\left(b_1^2-b_2^2+b_3^2\right)^2 \left(g_1^2-g_2^2+g_3^2\right)},\\
\al_{0}&=&2 b_1 (b_2 - b_3) \left[c_2 (d_2 g_1 - d_1 g_2)  + c_3 (d_1 g_3- d_3 g_1) +
    c_1 (d_1 g_1 - d_2 g_2 + d_3 g_3)\right]\crn
    &-&
 b_1^2 \left\{(c_3- c_2) [(d_2 - d_3) (g_2 + g_3)- d_1 g_1]+
    c_1 [(d_2- d_3) g_1 - d_1 (g_2 - g_3)]\right\} \crn
    &-& (b_2 - b_3)^2 \left\{c_1 [(d_2 + d_3) g_1 - d_1 (g_2 + g_3)]- (c_2+c_3) [(d_2 + d_3) (g_2 - g_3) - d_1 g_1]\right\}, \hspace{0.35cm}\\
\al_2&=& \frac{a_4}{d_1} + \frac{c_1 (a_3 d_1-a_4 b_1)}{b_1^2 g_1},\\
\al_3&=&\frac{\gamma_{0} a_3}{(b_1^2 - b_2^2 + b_3^2)^2 (g_1^2 - g_2^2 + g_3^2)}+\frac{a_4 d_1}{d_1^2-d_2^2+d_3^2} \crn
&+& \frac{a_4 \left\{-b_1 c_1 g_1 +
   b_1 (c_2 + c_3) (g_2 - g_3) + (b_2 + b_3) [(c_2- c_3)g_1 + (g_3-g_2) c_1]\right\}}{(b_1^2 - b_2^2 + b_3^2)^2 (g_1^2 - g_2^2 + g_3^2)},\\
\gamma_{0}&=&b_1^2 \left\{c_1 d_1 g_1 -
      c_1 (d_2 - d_3) (g_2 + g_3) + (c_2 + c_3) [(d_2 - d_3) g_1 - (g_2-g_3)d_1]\right\} \crn
      &+&2 b_1 \left\{-b_2 [(c_2 d_1+ c_1 d_2) g_1 - (c_1 d_1+ c_2 d_2+ c_3 d_3) g_2 +
         (c_3 d_2 + c_2 d_3) g_3] \right.\crn
         &+&\left. b_3 [(c_3 d_1 + c_1 d_3)g_1 - (c_3 d_2 + c_2 d_3) g_2 +(c_2 d_2+ c_3 d_3- c_1 d_1)g_3]\right\}\crn
         &+& (b_2^2 - b_3^2) \left\{c_1 [d_1 g_1 - (d_2 + d_3) (g_2 - g_3)] + (c_2 -
         c_3) [(d_2 + d_3) g_1 - d_1 (g_2 + g_3)]\right\}\hspace{-0.1 cm},\hspace{0.45 cm} \\
\beta_1&=&\frac{\beta_{0}}{\left(c_1^2-c_2^2+c_3^2\right)^2 \left(d_1^2-d_2^2+d_3^2\right)}, \\
\beta_{0}&=& (d_2^2-d_1^2 - d_3^2) c_1 a_3  +  (b_2^2-b_1^2 -b_3^2) g_1a_4 \crn
&+&\left\{b_1 d_1 g_1 -
    b_1 (d_2 + d_3) (g_2 - g_3) - (b_2 + b_3) [(d_2- d_3) g_1 - d_1 (g_2 - g_3)]\right\} a_3\crn
&+&\left\{b_1 c_1 d_1 + c_1(b_3-b_2) (d_2 + d_3) + (c_3-c_2)[(b_2  + b_3)d_1 +  b_1 (d_2 + d_3)]\right\}a_4,\\
\beta_2&=&\frac{(a_4 b_1 - a_3 d_1) (b_1 g_1-c_1 d_1)}{b_1 c_1 d_1 g_1}, \\
\beta_3&=&-\frac{a_4 g_1}{g_1^2-g_2^2+ g_3^2}-\frac{a_3 c_1}{c_1^2-c_2^2+c_3^2}\crn
&+&\frac{a4 \left\{b_1 c_1 d_1 +(c_2 + c_3)[(b_3- b_2)  d_1 +
    (d_2 - d_3) b_1] + (b_2 c_1 +b_3 c_1)(d_3-d_2)\right\}}{(c_1^2 -c_2^2 + c_3^2)^2 (d_1^2 -d_2^2 +d_3^2)}\crn
&+&\frac{a_3 \left\{(b_3 - b_2) [(d_2 + d_3) g_1 - d_1 (g_2 + g_3)] +
   b_1 [d_1 g_1 - (d_2 - d_3) (g_2 + g_3)]\right\}}{(b_1^2 -b_2^2 + b_3^2)^2 (g_1^2 -g_2^2 +g_3^2)}.\eea
\section{\label{Appenalbetaij} The explicit expressions of $\alpha_{13, 23, 14, 24}$ and $\beta_{13, 23, 14, 24}$}
  \vspace*{-0.25 cm}
\bea
&&\al_{13}=\left\{2 b_1 (b_2 - b_3) [(c_2 d_2- c_3 d_3) g_1 + (c_3 g_3- c_2 g_2)d_1  +
    c_1 (d_1 g_1 - d_2 g_2 + d_3 g_3)]\right. \crn
&&\left.-
b_1^2 [(c_2  - c_3) d_1 g_1 +(c_3- c_2) (d_2 - d_3) (g_2 + g_3) +
 (d_2- d_3) c_1g_1 + c_1d_1 (g_3-g_2)] \right. \crn
     &&\left.- (b_2 - b_3)^2 [(c_2 + c_3) d_1 g_1 + (d_2 + d_3) c_1 g_1 - (c_2+c_3) (d_2 + d_3) (g_2 - g_3) -  (g_2 + g_3)c_1 d_1]\right\}\crn
    &&/\left\{(b_1^2 - b_2^2 + b_3^2)^2 (g_1^2 - g_2^2 + g_3^2)\right\}, \\
&&\al_{23} = \frac{c_2 - c_3}{c_1^2 - c_2^2 + c_3^2}\crn
&&\hspace{0.725cm}+\, \frac{(b_3-b_2) d_1 g_1+ (b_2-b_3) (d_2 + d_3) (g_2 - g_3) + b_1 [(d_2  - d_3) g_1 +(g_3- g_2) d_1]}{(b_1^2 - b_2^2 + b_3^2)^2 (g_1^2 - g_2^2 + g_3^2)},  \\
&&\al_{14}=\frac{d_2-d_3}{d_1^2 - d_2^2 + d_3^2} \crn
&&\hspace{0.725cm}+\, \frac{(b_3-b_2) c_1 g_1 + (b_2-b_3) (c_2 + c_3) (g_2 - g_3) + b_1 [(c_2- c_3) g_1 +(g_3-g_2)c_1]}{(b_1^2 - b_2^2 + b_3^2)^2 (g_1^2 - g_2^2 + g_3^2)}, \\
&&\al_{24}=\frac{g_2-g_3}{g_1^2 - g_2^2 + g_3^2}\crn
&&\hspace{0.725cm}+\, \frac{(b_2- b_3) c_1 d_1 + (b_2 +b_3)(c_2 - c_3) (d_2 - d_3)  + b_1 [(c_3-c_2) d_1+ ( d_3-d_2) c_1]}{(c_1^2 - c_2^2 + c_3^2) (d_1^2 - d_2^2 + d_3^2)}, \\
&&\beta_{13}=\left\{2 b_1 (b_2 + b_3) [(c_2 d_2- c_3 d_3) g_1 + (c_3 g_3- c_2 g_2) d_1 +
    c_1 (d_1 g_1 - d_2 g_2 + d_3 g_3)]\right. \crn
&&\left.-
b_1^2 [(c_2 + c_3) d_1 g_1 -(c_2+c_3) (d_2 + d_3) (g_2 - g_3) +
 (d_2+ d_3) c_1g_1 - c_1d_1 (g_2+g_3)] \right. \crn
     &&\left.- (b_2 + b_3)^2 [(c_2 - c_3) d_1 g_1 + (d_2 + d_3) c_1 g_1 - (c_2+c_3) (d_2 + d_3) (g_2 - g_3) -  (g_2 + g_3)c_1 d_1]\right\}\crn
    &&/\left\{(b_1^2 - b_2^2 + b_3^2)^2 (g_1^2 - g_2^2 + g_3^2)\right\}, \\
&&\beta_{23} = \frac{c_2 +c_3}{c_1^2 - c_2^2 + c_3^2}\crn
&&\hspace{0.725cm}+\, \frac{-(b_2+b_3) d_1 g_1+ (b_2+b_3) (d_2 - d_3) (g_2 + g_3) + b_1 [(d_2  + d_3) g_1 -(g_2+ g_3) d_1]}{(b_1^2 - b_2^2 + b_3^2)^2 (g_1^2 - g_2^2 + g_3^2)},  \\
&&\beta_{14}=\frac{d_2+d_3}{d_1^2 - d_2^2 + d_3^2} \crn
&&\hspace{0.725cm}+\, \frac{-(b_2+b_3) c_1 g_1 + (b_2+b_3) (c_2 - c_3) (g_2+ g_3) + b_1 [(c_2+ c_3) g_1 -(g_2+g_3)c_1]}{(b_1^2 - b_2^2 + b_3^2)^2 (g_1^2 - g_2^2 + g_3^2)}, \\
&&\beta_{24}=\frac{g_2+g_3}{g_1^2 - g_2^2 + g_3^2}\crn
&&\hspace{0.725cm}+\, \frac{(b_2+ b_3) c_1 d_1 + (b_2-b_3)(c_2 + c_3) (d_2 + d_3)  - b_1 [(c_3+c_2) d_1+ ( d_2+d_3) c_1]}{(c_1^2 - c_2^2 + c_3^2) (d_1^2 - d_2^2 + d_3^2)}.\eea
\newpage
\section{Yukawa couplings prohibited by $U(1)_L, Z_4$ and $Z_2$}
\begin{table}[h]
\begin{center}
\vspace{-0.5 cm}
\caption{\label{U1Lprevent} Yukawa couplings prohibited by $U(1)_L$}
\vspace{0.25 cm}
 \begin{tabular}{|c|c|c|c|} \hline
$(\overline{\psi}_{L} N_{L}^C)_{\underline{3}_{a, s}} (\widetilde{H}\phi)_{\underline{3}}, (\overline{\psi}_{L} S_{L}^C)_{\underline{3}_{a, s}} (\widetilde{H}\phi)_{\underline{3}}, (\overline{N}_{L} \psi_{L}^C)_{\underline{3}_{a, s}} (\widetilde{H}\phi)_{\underline{3}}, (\overline{S}_{L} \psi_{L}^C)_{\underline{3}_{a, s}} (\widetilde{H}\phi)_{\underline{3}}, $ \\
$(\overline{N}_{L} N^C_{L})_{\underline{1}} \rho^{2}$, $(\overline{N}_{L} N^C_{L})_{\underline{1}} \rho^{*2}, (\overline{N}_{L} S^C_{L})_{\underline{1}} \rho^{2}, (\overline{N}_{L} S^C_{L})_{\underline{1}} \rho^{*2}, (\overline{S}_{L} N^C_{L})_{\underline{1}} \rho^{2}, (\overline{S}_{L} N^C_{L})_{\underline{1}} \rho^{*2}, $\\
 $(\overline{S}_{L} S^C_{L})_{\underline{1}} \rho^{2}, (\overline{S}_{L} S^C_{L})_{\underline{1}} \rho^{*2},  (\overline{N}_{L} N^C_{L})_{\underline{1}} (\phi\varphi)_{\underline{1}}, (\overline{N}_{L} N^C_{L})_{\underline{1}'} (\phi\varphi)_{\underline{1}''}, (\overline{N}_{L} N^C_{L})_{\underline{1}''} (\phi\varphi)_{\underline{1}'}, $\\
 $(\overline{N}_{L} N^C_{L})_{\underline{3}_{a, s}} (\phi\varphi)_{\underline{3}_{a, s}}, (\overline{N}_{L} S^C_{L})_{\underline{1}} (\phi\varphi)_{\underline{1}}$, $ (\overline{N}_{L} S^C_{L})_{\underline{1}'} (\phi\varphi)_{\underline{1}''}, (\overline{N}_{L} S^C_{L})_{\underline{1}''} (\phi\varphi)_{\underline{1}'},$ \\
$ (\overline{N}_{L} S^C_{L})_{\underline{3}_{a, s}} (\phi\varphi)_{\underline{3}_{a, s}}$,
$(\overline{S}_{L} N^C_{L})_{\underline{1}} (\phi\varphi)_{\underline{1}}, (\overline{S}_{L} N^C_{L})_{\underline{1}'} (\phi\varphi)_{\underline{1}''}$, $ (\overline{S}_{L} N^C_{L})_{\underline{1}''} (\phi\varphi)_{\underline{1}'}$,\\
$(\overline{S}_{L} N^C_{L})_{\underline{3}_{a, s}} (\phi\varphi)_{\underline{3}_{a, s}}$,
$(\overline{S}_{L} S^C_{L})_{\underline{1}} (\phi\varphi)_{\underline{1}}, (\overline{S}_{L} S^C_{L})_{\underline{1}'} (\phi\varphi)_{\underline{1}''}, (\overline{S}_{L} S^C_{L})_{\underline{1}''} (\phi\varphi)_{\underline{1}'}, $\\
$ (\overline{S}_{L} S^C_{L})_{\underline{3}_{a, s}} (\phi\varphi)_{\underline{3}_{a, s}}$,
$(\overline{N}_{L} N^C_{L})_{\underline{3}_{a, s}} (\phi\chi)_{\underline{3}},  (\overline{N}_{L} S^C_{L})_{\underline{3}_{a, s}} (\phi\chi)_{\underline{3}}, (\overline{S}_{L} N^C_{L})_{\underline{3}_{a, s}} (\phi\chi)_{\underline{3}},$ \\
 $ (\overline{S}_{L} S^C_{L})_{\underline{3}_{a, s}} (\phi\chi)_{\underline{3}}$, $(\overline{N}_{R} N^C_{R})_{\underline{3}_{a, s}} (\phi\chi)_{\underline{3}}$,
$(\overline{N}_{R} S^C_{R})_{\underline{3}_{a, s}} (\phi\chi)_{\underline{3}}, (\overline{S}_{R} N^C_{R})_{\underline{3}_{a, s}} (\phi\chi)_{\underline{3}},$\\
$(\overline{S}_{R} S^C_{R})_{\underline{3}_{a, s}} (\phi\chi)_{\underline{3}}, (\overline{\nu}^C_{R} \nu_{R})_{\underline{1}} (\phi^2)_{\underline{1}}, (\overline{\nu}^C_{R} \nu_{R})_{\underline{1}'} (\phi^2)_{\underline{1}''}, (\overline{\nu}^C_{R} \nu_{R})_{\underline{1}''} (\phi^2)_{\underline{1}'}, $\\
$(\overline{\nu}^C_{R} \nu_{R})_{\underline{3}_{s,a}} (\phi^2)_{\underline{3}_{s,a}}, (\overline{\nu}^C_{R} \nu_{R})_{\underline{1}} (\varphi^2)_{\underline{1}}, (\overline{\nu}^C_{R} \nu_{R})_{\underline{1}'} (\varphi^2)_{\underline{1}''}, (\overline{\nu}^C_{R} \nu_{R})_{\underline{1}''} (\varphi^2)_{\underline{1}'},$ \\
$(\overline{\nu}^C_{R} \nu_{R})_{\underline{3}_{s,a}} (\varphi^2)_{\underline{3}_{s,a}}, (\overline{\nu}^C_{R} \nu_{R})_{\underline{1}} (\chi^2)_{\underline{1}}, (\overline{\nu}^C_{R} \nu_{R})_{\underline{3}_{s, a}} (\varphi \chi)_{\underline{3}}, (\overline{\nu}^C_{R} N_{R})_{\underline{3}_{s,a}} (\phi \rho)_{\underline{3}},$ \\
$ (\overline{\nu}^C_{R} S_{R})_{\underline{3}_{s,a}} (\phi \rho)_{\underline{3}},
(\overline{N}^C_{R} \nu_{R})_{\underline{3}_{s,a}} (\phi \rho)_{\underline{3}}, (\overline{S}^C_{R} \nu_{R})_{\underline{3}_{s,a}} (\phi \rho)_{\underline{3}}, (\overline{\nu}^C_{R} N_{R})_{\underline{3}_{s,a}} (\phi^* \rho)_{\underline{3}}, $\\
$(\overline{\nu}^C_{R} S_{R})_{\underline{3}_{s,a}} (\varphi\rho^*)_{\underline{3}},
(\overline{N}^C_{R} \nu_{R})_{\underline{3}_{s,a}} (\varphi\rho^*)_{\underline{3}}, $
$(\overline{S}^C_{R} \nu_{R})_{\underline{3}_{s,a}} (\varphi\rho^*)_{\underline{3}},
(\overline{\nu}^C_{R} N_{R})_{\underline{1}} (\chi \rho^*)_{\underline{1}}$,\\
$(\overline{\nu}^C_{R} S_{R})_{\underline{1}} (\chi \rho^*)_{\underline{1}},
(\overline{N}^C_{R} \nu_{R})_{\underline{1}} (\chi \rho^*)_{\underline{1}}$,
$(\overline{S}^C_{R} \nu_{R})_{\underline{1}} (\chi \rho^*)_{\underline{1}},$\\ \hline
\end{tabular}
\vspace{-0.25 cm}
\end{center}
\end{table}
\begin{table}[h]
\begin{center}
\caption{\label{Z4prevent} Yukawa couplings prohibited by $Z_4$}
\vspace{0.25 cm}
 \begin{tabular}{|c|c|c|c|} \hline
$(\overline{\psi}_L l_{iR})_{\underline{3}} (H\varphi)_{\underline{3}}$,
$(\overline{\psi}_{L} \nu_{R})_{\underline{3}_{a,s}} (\widetilde{H}\phi)_{\underline{3}}$,
$(\overline{\psi}_{L} \nu_{R})_{\underline{3}_{a,s}} (\widetilde{H}\varphi)_{\underline{3}}, (\overline{N}_{L} \nu_{R})_{\underline{1}} \rho^*, $ \\
$(\overline{S}_{L} \nu_{R})_{\underline{1}} \rho^*, (\overline{N}_{L} \nu_{R})_{\underline{1}} (\phi^2)_{\underline{1}}, (\overline{N}_{L} \nu_{R})_{\underline{1}'} (\phi^2)_{\underline{1}''}, (\overline{N}_{L} \nu_{R})_{\underline{1}''} (\phi^2)_{\underline{1}'}, $\\
$(\overline{N}_{L} \nu_{R})_{\underline{3}_{a,s}} (\phi^2)_{\underline{3}_{a,s}}, (\overline{S}_{L} \nu_{R})_{\underline{1}} (\phi^2)_{\underline{1}}$, $(\overline{S}_{L} \nu_{R})_{\underline{1}'} (\phi^2)_{\underline{1}''}, (\overline{S}_{L} \nu_{R})_{\underline{1}''} (\phi^2)_{\underline{1}'}$, \\
$(\overline{S}_{L} \nu_{R})_{\underline{3}_{a,s}} (\phi^2)_{\underline{3}_{a,s}}, (\overline{N}_{L} \nu_{R})_{\underline{1}} (\varphi^2)_{\underline{1}}, (\overline{N}_{L} \nu_{R})_{\underline{1}'} (\varphi^2)_{\underline{1}''}, (\overline{N}_{L} \nu_{R})_{\underline{1}''} (\varphi^2)_{\underline{1}'},$ \\
$ (\overline{S}_{L} \nu_{R})_{\underline{1}} (\varphi^2)_{\underline{1}}, (\overline{S}_{L} \nu_{R})_{\underline{1}'} (\varphi^2)_{\underline{1}''}, (\overline{S}_{L} \nu_{R})_{\underline{1}''} (\varphi^2)_{\underline{1}'}, (\overline{N}_{L} \nu_{R})_{\underline{1}} \chi^2, (\overline{N}_{L} \nu_{R})_{\underline{1}} \rho^2,$\\
$  (\overline{S}_{L} \nu_{R})_{\underline{1}} \chi^2, (\overline{S}_{L} \nu_{R})_{\underline{1}} \rho^2, (\overline{N}_{L} \nu_{R})_{\underline{1}} (\phi \varphi)_{\underline{1}},
(\overline{N}_{L} \nu_{R})_{\underline{1}'} (\phi \varphi)_{\underline{1}''}, (\overline{N}_{L} \nu_{R})_{\underline{1}''} (\phi \varphi)_{\underline{1}'},$ \\

$ (\overline{N}_{L} \nu_{R})_{\underline{3}_{a,s}} (\phi \varphi)_{\underline{3}_{a,s}}, $
$ (\overline{S}_{L} \nu_{R})_{\underline{1}} (\phi \varphi)_{\underline{1}}, (\overline{S}_{L} \nu_{R})_{\underline{1}'} (\phi \varphi)_{\underline{1}''}, (\overline{S}_{L} \nu_{R})_{\underline{1}''} (\phi \varphi)_{\underline{1}'},$ \\
$(\overline{S}_{L} \nu_{R})_{\underline{3}_{a,s}} (\phi \varphi)_{\underline{3}_{a,s}}, (\overline{N}_{L} \nu_{R})_{\underline{3}_s} (\phi\chi)_{\underline{3}}, (\overline{N}_{L} \nu_{R})_{\underline{3}_a} (\phi\chi)_{\underline{3}}, (\overline{N}_{L} \nu_{R})_{\underline{3}_s} (\varphi \chi)_{\underline{3}},$\\
$ (\overline{N}_{L} \nu_{R})_{\underline{3}_a} (\varphi \chi)_{\underline{3}}, (\overline{S}_{L} \nu_{R})_{\underline{3}_s} (\phi\chi)_{\underline{3}}, (\overline{S}_{L} \nu_{R})_{\underline{3}_a} (\phi\chi)_{\underline{3}}, (\overline{S}_{L} \nu_{R})_{\underline{3}_s} (\varphi \chi)_{\underline{3}},$ \\

$(\overline{S}_{L} \nu_{R})_{\underline{3}_a} (\varphi \chi)_{\underline{3}}, (\overline{N}_{L} N_{R})_{\underline{3}_{s,a}} (\phi \rho)_{\underline{3}},
(\overline{N}_{L} S_{R})_{\underline{3}_{s,a}} (\phi \rho)_{\underline{3}}, (\overline{S}_{L} N_{R})_{\underline{3}_{s,a}} (\phi \rho)_{\underline{3}},$ \\
$ (\overline{S}_{L} S_{R})_{\underline{3}_{s,a}} (\phi \rho)_{\underline{3}} , $

$ (\overline{N}_{L} N_{R})_{\underline{3}_{s,a}} (\phi \rho^*)_{\underline{3}},
(\overline{N}_{L} S_{R})_{\underline{3}_{s,a}} (\phi \rho^*)_{\underline{3}}, (\overline{S}_{L} N_{R})_{\underline{3}_{s,a}} (\phi \rho^*)_{\underline{3}},$  \\
$
(\overline{S}_{L} S_{R})_{\underline{3}_{s,a}} (\phi \rho^*)_{\underline{3}}, $

$(\overline{N}_{L} N_{R})_{\underline{3}_{s,a}} (\varphi \rho)_{\underline{3}}, (\overline{N}_{L} N_{R})_{\underline{3}_{s,a}} (\varphi \rho^*)_{\underline{3}},
(\overline{N}_{L} S_{R})_{\underline{3}_{s,a}} (\varphi \rho)_{\underline{3}},$ \\
$ (\overline{N}_{L} S_{R})_{\underline{3}_{s,a}} (\varphi \rho^*)_{\underline{3}}, (\overline{S}_{L} N_{R})_{\underline{3}_{s,a}} (\varphi \rho)_{\underline{3}}, (\overline{S}_{L} N_{R})_{\underline{3}_{s,a}} (\varphi \rho^*)_{\underline{3}},
(\overline{S}_{L} S_{R})_{\underline{3}_{s,a}} (\varphi \rho)_{\underline{3}}, $ \\

$(\overline{S}_{L} S_{R})_{\underline{3}_{s,a}} (\varphi \rho^*)_{\underline{3}}, (\overline{N}_{L} N_{R})_{\underline{1}} (\chi \rho)_{\underline{1}}, (\overline{N}_{L} N_{R})_{\underline{1}} (\chi \rho^*)_{\underline{1}},
(\overline{N}_{L} S_{R})_{\underline{1}} (\chi \rho)_{\underline{1}},  $ \\
$(\overline{N}_{L} S_{R})_{\underline{1}} (\chi \rho^*)_{\underline{1}}, (\overline{S}_{L} N_{R})_{\underline{1}} (\chi \rho)_{\underline{1}}, (\overline{S}_{L} N_{R})_{\underline{1}} (\chi \rho^*)_{\underline{1}},
(\overline{S}_{L} S_{R})_{\underline{1}} (\chi \rho)_{\underline{1}},  (\overline{S}_{L} S_{R})_{\underline{1}} (\chi \rho^*)_{\underline{1}},$\\
$ (\overline{\nu}^C_{R} \nu_{R})_{\underline{1}} \rho. $ \\ \hline
\end{tabular}
\vspace{-1.25 cm}
\end{center}
\end{table}
\newpage
\begin{table}[h]
\begin{center}
\vspace{-0.25 cm}
\caption{\label{Z2prevent} Yukawa couplings prohibited by $Z_2$}
\vspace{0.25 cm}
 \begin{tabular}{|c|c|c|c|} \hline
$(\overline{\psi}_{L} N_{R})_{\underline{3}_{a,s}} (\widetilde{H}\phi)_{\underline{3}}, (\overline{\psi}_{L} N_{R})_{\underline{3}_{a,s}} (\widetilde{H}\phi^*)_{\underline{3}}, (\overline{\psi}_{L} S_{R})_{\underline{3}_{a,s}} (\widetilde{H}\phi)_{\underline{3}}, $\\

$(\overline{\psi}_{L} S_{R})_{\underline{3}_{a,s}} (\widetilde{H}\phi^*)_{\underline{3}}, (\overline{N}_{L} \nu_{R})_{\underline{3}_s} (\phi \rho)_{\underline{3}}, (\overline{N}_{L} \nu_{R})_{\underline{3}_a} (\phi \rho)_{\underline{3}}, (\overline{S}_{L} \nu_{R})_{\underline{3}_s} (\phi \rho)_{\underline{3}},$\\

$ (\overline{S}_{L} \nu_{R})_{\underline{3}_a} (\phi \rho)_{\underline{3}}, (\overline{N}_{L} \nu_{R})_{\underline{3}_s} (\phi^* \rho)_{\underline{3}}, (\overline{N}_{L} \nu_{R})_{\underline{3}_a} (\phi^* \rho)_{\underline{3}}, (\overline{S}_{L} \nu_{R})_{\underline{3}_s} (\phi^* \rho)_{\underline{3}}, $\\

$(\overline{S}_{L} \nu_{R})_{\underline{3}_a} (\phi^* \rho)_{\underline{3}}, (\overline{N}_{L} \nu_{R})_{\underline{3}_s} (\varphi \rho^*)_{\underline{3}}, (\overline{N}_{L} \nu_{R})_{\underline{3}_a} (\varphi \rho^*)_{\underline{3}}, (\overline{S}_{L} \nu_{R})_{\underline{3}_s} (\varphi \rho^*)_{\underline{3}}, $\\

$(\overline{S}_{L} \nu_{R})_{\underline{3}_a} (\varphi \rho^*)_{\underline{3}}, (\overline{N}_{L} \nu_{R})_{\underline{3}_s} (\chi \rho^*)_{\underline{3}}, (\overline{N}_{L} \nu_{R})_{\underline{3}_a} (\chi \rho^*)_{\underline{3}}, (\overline{S}_{L} \nu_{R})_{\underline{3}_s} (\chi \rho^*)_{\underline{3}},$\\
$ (\overline{S}_{L} \nu_{R})_{\underline{3}_a} (\chi \rho^*)_{\underline{3}}, (\overline{N}_{L} N_{R})_{\underline{1}} (\rho^{2})_{\underline{1}}, (\overline{N}_{L} N_{R})_{\underline{1}} (\rho^{*2})_{\underline{1}},
(\overline{N}_{L} S_{R})_{\underline{1}} (\rho^{2})_{\underline{1}}, (\overline{N}_{L} S_{R})_{\underline{1}} (\rho^{*2})_{\underline{1}},$\\
$(\overline{S}_{L} N_{R})_{\underline{1}} (\rho^{2})_{\underline{1}}, (\overline{S}_{L} N_{R})_{\underline{1}} (\rho^{*2})_{\underline{1}},
(\overline{S}_{L} S_{R})_{\underline{1}} (\rho^{2})_{\underline{1}}, (\overline{S}_{L} S_{R})_{\underline{1}} (\rho^{*2})_{\underline{1}}, $\\

$(\overline{N}_{L} N_{R})_{\underline{1}} (\phi \varphi)_{\underline{1}}, (\overline{N}_{L} N_{R})_{\underline{1}'} (\phi \varphi)_{\underline{1}''}, (\overline{N}_{L} N_{R})_{\underline{1}''} (\phi \varphi)_{\underline{1}'}, (\overline{N}_{L} N_{R})_{\underline{3}_{s,a}} (\phi \varphi)_{\underline{3}_{s,a}}, $\\
$(\overline{N}_{L} N_{R})_{\underline{1}} (\phi^* \varphi)_{\underline{1}}, (\overline{N}_{L} N_{R})_{\underline{1}'} (\phi^* \varphi)_{\underline{1}''}, (\overline{N}_{L} N_{R})_{\underline{1}''} (\phi^* \varphi)_{\underline{1}'}, (\overline{N}_{L} N_{R})_{\underline{3}_{s,a}} (\phi^* \varphi)_{\underline{3}_{s,a}}, $\\

$(\overline{N}_{L} S_{R})_{\underline{1}} (\phi \varphi)_{\underline{1}}, (\overline{N}_{L} S_{R})_{\underline{1}'} (\phi \varphi)_{\underline{1}''}, (\overline{N}_{L} S_{R})_{\underline{1}''} (\phi \varphi)_{\underline{1}'}, (\overline{N}_{L} S_{R})_{\underline{3}_{s,a}} (\phi \varphi)_{\underline{3}_{s,a}}, $\\
$(\overline{N}_{L} S_{R})_{\underline{1}} (\phi^* \varphi)_{\underline{1}}, (\overline{N}_{L} S_{R})_{\underline{1}'} (\phi^* \varphi)_{\underline{1}''}, (\overline{N}_{L} S_{R})_{\underline{1}''} (\phi^* \varphi)_{\underline{1}'}, (\overline{N}_{L} S_{R})_{\underline{3}_{s,a}} (\phi^* \varphi)_{\underline{3}_{s,a}}, $\\

$(\overline{S}_{L} N_{R})_{\underline{1}} (\phi \varphi)_{\underline{1}}, (\overline{S}_{L} N_{R})_{\underline{1}'} (\phi \varphi)_{\underline{1}''}, (\overline{S}_{L} N_{R})_{\underline{1}''} (\phi \varphi)_{\underline{1}'}, (\overline{S}_{L} N_{R})_{\underline{3}_{s,a}} (\phi \varphi)_{\underline{3}_{s,a}}, $\\
$(\overline{S}_{L} N_{R})_{\underline{1}} (\phi^* \varphi)_{\underline{1}}, (\overline{S}_{L} N_{R})_{\underline{1}'} (\phi^* \varphi)_{\underline{1}''}, (\overline{S}_{L} N_{R})_{\underline{1}''} (\phi^* \varphi)_{\underline{1}'}, (\overline{S}_{L} N_{R})_{\underline{3}_{s,a}} (\phi^* \varphi)_{\underline{3}_{s,a}}, $\\

$(\overline{S}_{L} S_{R})_{\underline{1}} (\phi \varphi)_{\underline{1}}, (\overline{S}_{L} S_{R})_{\underline{1}'} (\phi \varphi)_{\underline{1}''}, (\overline{S}_{L} S_{R})_{\underline{1}''} (\phi \varphi)_{\underline{1}'}, (\overline{S}_{L} S_{R})_{\underline{3}_{s,a}} (\phi \varphi)_{\underline{3}_{s,a}}, $\\
$(\overline{S}_{L} S_{R})_{\underline{1}} (\phi^* \varphi)_{\underline{1}}, (\overline{S}_{L} S_{R})_{\underline{1}'} (\phi^* \varphi)_{\underline{1}''}, (\overline{S}_{L} S_{R})_{\underline{1}''} (\phi^* \varphi)_{\underline{1}'}, (\overline{S}_{L} S_{R})_{\underline{3}_{s,a}} (\phi^* \varphi)_{\underline{3}_{s,a}}, $\\
$(\overline{N}_{L} N_{R})_{\underline{3}_{s,a}} (\varphi \chi)_{\underline{3}},
(\overline{N}_{L} S_{R})_{\underline{3}_{s,a}} (\varphi \chi)_{\underline{3}},
(\overline{S}_{L} N_{R})_{\underline{3}_{s,a}} (\varphi \chi)_{\underline{3}},
(\overline{S}_{L} S_{R})_{\underline{3}_{s,a}} (\varphi \chi)_{\underline{3}}, $\\  \hline
\end{tabular}
\end{center}
\end{table}
\vspace{-0.5 cm}
\section{\label{anomaly}The anomaly free of the model}
For convenience, let us list the quantum numbers of quarks fields in Table \ref{quanyumnumber}.
\begin{table}[h]
\caption[]{\label{quanyumnumber} $SU(3)_C$, $SU(2)_L$, $U(1)_Y$ and $U(1)_L$ quantum numbers of quarks fields}
\vspace{-0.25 cm}
\bc
\begin{tabular}{|c|c|c|c|c|c|ccc}
\hline
& $SU(3)_C$ & $SU(2)_L$& $U(1)_Y$& $U(1)_L$  \\ \hline
\hline
$q_L =\left(u_{L} \hs
d_{L}\right)^T$ & $3$ & $2$ & $\frac{1}{6}$ & $-\frac{1}{3}$   \\ \hline
$u_{R}$ & $3$ & $1$ & $\frac{2}{3}$ & $-\frac{1}{3}$  \\ \hline
$d_{R}$ & $3$ & $1$ & $-\frac{1}{3}$ & $-\frac{1}{3}$  \\ \hline
\end{tabular}
\ec
\end{table} \\
In the considered model, all anomalies are cancelled within each generation, because of
\bea
&&[SU(3)_C]^2 U(1)_Y \sim  \sum_{\mathrm{quarks}} (Y_{q_L}-Y_{q_R}) = 3(2Y_{q_L} - Y_{u} - Y_{d})\crn
&&\hspace{3 cm}= 3\left[2\times \frac{1}{6}-\left(\frac{2}{3}-\frac{1}{3}\right)\right]=0, \\
&&[SU(3)_C]^2 U(1)_{L} \sim \sum_{\mathrm{quarks}} (L_{q_L}-L_{q_R}) =  3(2L_{q_L} - L_{u} - L_{d}) \crn
&&\hspace{3 cm} = 3\left[2\left(-\frac{1}{3}\right)-\left(-\frac{1}{3}-\frac{1}{3}\right)\right]=0, \\
&&[SU(2)_L]^2 U(1)_Y\sim \sum_{\mathrm{doublets}} Y_{f_L}= Y_{\psi_L}+3Y_{q_L} = -\frac{1}{2}+3\times \frac{1}{6} =0, \eea
\bea
&&[SU(2)_L]^2 U(1)_L \sim \sum_{\mathrm{doublets}} L_{f_L}=L_{\psi_L}+3L_{q_L}= 1+3\left(-\frac{1}{3}\right) = 0,\\
&&[\mathrm{Gravity}]^2U(1)_Y \sim \sum_{\mathrm{fermions}}(Y_{f_L}-Y_{f_R})=2Y_{\psi_{L}}+Y_{N_{L}}+Y_{S_{L}}+3\times 2Y_{q_{L}}-Y_{l_{R}} \crn
&&\hspace{3 cm}-Y_{\nu_{R}}-Y_{N_{R}}-Y_{S_{R}}-3Y_{u_{R}}-3Y_{d_{R}}
=2\left(-\frac{1}{2}\right)+0+0\crn
&&\hspace{3 cm}+3\times 2 \times \frac{1}{6}-(-1)-0-0-0-3\times \frac{2}{3}-3\left(-\frac{1}{3}\right)=0, \\
&&[\mathrm{Gravity}]^2U(1)_L \sim\sum_{\mathrm{fermions}}(L_{f_L}-L_{f_R})= 2L_{\psi_{L}}+L_{N_{L}}+L_{S_{L}}+3\times 2L_{q_{L}}\crn
&&\hspace{3 cm}-L_{l_{R}}-L_{\nu_{R}}-L_{N_{R}}-L_{S_{R}}-3L_{u_{R}}-3L_{d_{R}}
=2\times 1+1+1\crn
&&\hspace{3 cm}+3\times 2 \left(-\frac{1}{3}\right) -1-1-1-1-3\left(-\frac{1}{3}\right)-3\left(-\frac{1}{3}\right)=0, \\
&&[U(1)_Y]^2U(1)_L\sim \sum_{\mathrm{fermions}}(Y^2_{f_L} L_{f_L}-Y^2_{f_R}L_{f_R}) =2 Y^2_{\psi_{L}} L_{\psi_{L}}+ Y^2_{N_{L}}L_{N_{L}}+Y^2_{S_{L}}L_{S_{L}}\crn
&&\hspace{2.75 cm} + 3\times 2Y^2_{q_{L}}L_{q_{L}} -Y^2_{l_{R}}L_{l_{R}}-Y^2_{\nu_{R}}L_{\nu_{R}}-Y^2_{N_{R}}L_{N_{R}}-Y^2_{S_{R}}L_{S_{R}}\crn
&&\hspace{2.75 cm}-3Y^2_{u_{R}}L_{u_{R}}-3Y^2_{d_{R}}L_{d_{R}}=2 \left(-\frac{1}{2}\right)^2\times 1 + 0^2\times 1+0^2\times 1\crn
&&\hspace{2.75 cm}+3\times 2 \left(\frac{1}{6}\right)^2\left(-\frac{1}{3}\right)-(-1)^2\times 1-0^2\times 1-0^2\times 1-0^2\times 1 \crn
&&\hspace{2.75 cm}-3\left(\frac{2}{3}\right)^2 \left(-\frac{1}{3}\right)-3\left(-\frac{1}{3}\right)^2 \left(-\frac{1}{3}\right)=0, \\
&&[U(1)_L]^2U(1)_Y\sim \sum_{\mathrm{fermions}}(L^2_{f_L} Y_{f_L}-L^2_{f_R}Y_{f_R})
=2 L^2_{\psi_{L}} Y_{\psi_{L}}+ L^2_{N_{L}}Y_{N_{L}}+L^2_{S_{L}}Y_{S_{L}}\crn
&&\hspace{2.75 cm}+3\times 2L^2_{q_{L}}Y_{q_{L}}-L^2_{l_{R}}Y_{l_{R}}-L^2_{\nu_{R}}Y_{\nu_{R}}-L^2_{N_{R}}Y_{N_{R}}-L^2_{S_{R}}Y_{S_{R}}\crn
&&\hspace{2.75 cm}-3L^2_{u_{R}}Y_{u_{R}}-3L^2_{d_{R}}Y_{d_{R}}
=2\times 1^2\left(-\frac{1}{2}\right) + 1^2\times 0+1^2\times 0\crn
&&\hspace{2.75 cm}+3\times 2 \left(-\frac{1}{3}\right)^2\left(\frac{1}{6}\right)-1^2(-1)-1^2\times 0-1^2\times 0-1^2\times 0\crn
&&\hspace{2.75 cm}-3 \left(-\frac{1}{3}\right)^2\left(\frac{2}{3}\right)-3\left(-\frac{1}{3}\right)^2\left(-\frac{1}{3}\right)=0, \\
&&[U(1)_Y]^3 \sim \sum_{\mathrm{fermions}}(Y^3_{f_L}-Y^3_{f_R})=2 Y^3_{\psi_{L}}+ Y^3_{N_{L}}+Y^3_{S_{L}}+3\times 2 Y^3_{q_{L}}\crn
&&\hspace{1.65 cm} - Y^3_{l_{R}}-Y^3_{\nu_{R}}-Y^3_{N_{R}}-Y^3_{S_{R}}-3Y^3_{u_{R}}-3Y^3_{d_{R}}\crn
&&\hspace{1.65 cm}=2\left(-\frac{1}{2}\right)^3 + 0^3+0^3+3\times 2 \left(\frac{1}{6}\right)^3
-(-1)^3-0^3\crn
&&\hspace{1.65 cm}-0^3-0^3-3 \left(\frac{2}{3}\right)^3-3\left(-\frac{1}{3}\right)^3=0,
\eea
\bea
&&[U(1)_L]^3\sim \sum_{\mathrm{fermions}}(L^3_{f_L}-L^3_{f_R})=2 L^3_{\psi_{L}}+ L^3_{N_{L}}+L^3_{S_{L}}+3\times 2 L^3_{q_{L}}\crn
&&\hspace{1.60 cm}-L^3_{l_{R}}-L^3_{\nu_{R}}-L^3_{N_{R}}-L^3_{S_{R}}-3L^3_{u_{R}}-3L^3_{d_{R}}\crn
&&\hspace{1.65 cm}=2\times 1^3 + 1^3+1^3+3\times 2 \left(-\frac{1}{3}\right)^3-1^3-1^3\crn
&&\hspace{1.65 cm}-1^3-1^3-3 \left(-\frac{1}{3}\right)^3-3\left(-\frac{1}{3}\right)^3=0. \hs\hs\hs
\eea


\begin{thebibliography}{99}
\bibitem{Salas2021} P. F. de Salas \emph{et al.}\, \emph{J. High Energ. Phys.} \textbf{2021} 71,  (2021)
 arXiv: 2006.11237 [hep-ph]. DOI: 10.1007/JHEP02(2021)071.

\bibitem{Katrin21PRL} M. Aker \emph{et al.} (KATRIN Collaboration) 
    \emph{Phys. Rev. Lett.} \textbf{123} \, 221802 (2019). 
\bibitem{Katrin21PRD} M. Aker \emph{et al.} (KATRIN Collaboration) 
    \emph{Phys. Rev. D} \textbf{104}  012005 (2021). 
\bibitem{Katrin21arX} M. Aker \emph{et al.} (KATRIN Collaboration) 
\emph{Nat. Phys.} \textbf{18}  160 (2022). 

\bibitem{seesaw123} Y. Cai, T. Han, T. Li, R. Ruiz 
\emph{Front.in Phys.} \textbf{6}  40 (2018). 


\bibitem{seesaw1} J. Schechter and J. W. F. Valle 
\emph{Phys. Rev. D} \textbf{22}  2227 (1980). 
\bibitem{seesaw2} J. Schechter and J. W. F. Valle 
\emph{Phys. Rev. D} \textbf{25}  774 (1982). 
\bibitem{seesaw3} R. N. Mohapatra and J. W. F. Valle 
\emph{Phys. Rev. D} \textbf{34}  1642 (1986). 
\bibitem{seesaw4} R. N. Mohapatra 
\emph{Phys. Rev. Lett.} \textbf{56}  561 (1986). 
\bibitem{seesaw5} J. Bernabeu, A. Santamaria, J. Vidal, A. Mendez and J. W. F. Valle 
\emph{Phys. Lett. B} \textbf{187} 303 (1987). 




\bibitem{A41} E. Ma and G. Rajasekaran 
\emph{Phys. Rev. D} \textbf{64} 113012 (2001). 
\bibitem{A42} E. Ma 
\emph{Mod. Phys. Lett. A} \textbf{17} 289 (2002). 
\bibitem{A43} E. Ma 
\emph{Mod. Phys. Lett. A} \textbf{17} 627 (2002). 

\bibitem{A44} K. S. Babu, E. Ma and J. W. F. Valle 
    \emph{Phys. Lett. B} \textbf{552} 207 (2003). 
\bibitem{A46} X. G. He, Y. Y. Keum and R. R. Volkas 
    \emph{J. High Energ. Phys.} \textbf{0604}  039 (2006). 
\bibitem{A47} G. Altarelli, F. Feruglio and Y. Lin 
\emph{Nucl. Phys. B} \textbf{775}  31(2007). 
\bibitem{A48} S. Morisi, M. Picariello and E. Torrente-Lujan 
\emph{Phys. Rev. D} \textbf{75} 075015 (2007). 
\bibitem{A49} F. Bazzocchi, S. Kaneko and S. Morisi 
\emph{J. High Energ. Phys.} \textbf{0803} 063 (2008). 
\bibitem{A410} F. Bazzocchi, M. Frigerio and S. Morisi 
\emph{Phys. Rev. D} \textbf{78} 116018 (2008). 
\bibitem{A411} G. Altarelli, F. Feruglio and C. Hagedorn 
\emph{J. High Energ. Phys.} \textbf{0803}  052 (2008). 
\bibitem{A412} M. Hirsch, S. Morisi and J. W. F. Valle 
\emph{Phys. Rev. D} \textbf{78} 093007 (2008). 
\bibitem{A413} E. Ma 
\emph{Phys. Lett. B} \textbf{671}  366 (2009). 
\bibitem{A414} G. Altarelli and D. Meloni 
\emph{J. Phys. G} \textbf{36}  085005 (2009). 
\bibitem{A415} Y. Lin 
\emph{Nucl. Phys. B} \textbf{813}  91 (2009).
\bibitem{A416} Y. H. Ahn and C. S. Chen 
\emph{Phys. Rev. D} \textbf{81}  105013 (2010).
\bibitem{A417} J. Barry and W. Rodejohanny 
\emph{Phys. Rev. D} \textbf{81}  093002(2010). 
\bibitem{A418} P. V. Dong, L. T. Hue, H. N. Long and D. V. Soa 
\emph{Phys. Rev. D} \textbf{81}  053004 (2010). 
\bibitem{A419} G. J. Ding and D. Meloni 
\emph{Nucl. Phys. B} \textbf{855}  21 (2012). 


\bibitem{A420Ishimori10} H. Ishimori,  T. Kobayashi, H. Ohki, H. Okada, Y. Shimizu and M. Tanimoto 
    \emph{Prog.Theor.Phys.Suppl.} \textbf{183}  1 (2010). 
\bibitem{A421VL2015} V. V. Vien and H. N. Long 
    \emph{Int. J. Mod. Phys. A} \textbf{30}  1550117 (2015). 
\bibitem{A422} T. Phong Nguyen, L. T. Hue, D. T. Si and T. T. Thuc 
\emph{Prog. Theor. Exp. Phys.} \textbf{2020} 033B04 (2020). 
\bibitem{A423} Gui-Jun Ding, Jun-Nan Lu and J. W. F. Valle 
\emph{Phys. Lett. B} \textbf{815}  136122 (2021). 
\bibitem{A424} V. V. Vien 
\emph{Mod. Phys. Lett. A} \textbf{35}  2050311 (2020). 
\bibitem{A425} V. V. Vien \emph{J. Phys. G: Nucl. Part. Phys.} \textbf{49} 085001 (2022).




\bibitem{LseesawA41} M. Hirsch, S. Morisi and J. W. F. Valle 
\emph{Phys. Lett. B} \textbf{679}  454 (2009). 
\bibitem{LseesawA45} D. Boraha and B. Karmakar 
\emph{Phys. Lett. B} \textbf{789} 59 (2019). 
\bibitem{LseesawA46} V. V. Vien, H. N. Long and A. E. Cárcamo Hernández 
 \emph{Phys.Lett.B} \textbf{798} 134979 (2019). 
\bibitem{LseesawS32} A. E. C\'arcamo Hern\'andez, R. Martínez and F. Ochoa 
    \emph{Eur. Phys.J. C} \textbf{76}  634 (2016). 


\bibitem{LseesawA43} M. Sruthilaya, R. Mohanta and S. Patra 
\emph{Eur. Phys. J. C} \textbf{78}  719 (2018). 
\bibitem{LseesawAD274} A.~E.~ C\'arcamo Hern\'andez, S.~Kovalenko, J.~W.~F.~Valle and C.~A.~Vaquera-Araujo 
    \emph{J. High Energ. Phys.} \textbf{1707} 118 (2017). 


\bibitem{LseesawD277}  A.~E.~ C\'arcamo Hern\'andez, S.~Kovalenko, J.~W.~F.~Valle and C.~A.~Vaquera-Araujo 
    \emph{J. High Energ. Phys.} \textbf{1902} 065 (2019). 
\bibitem{LseesawS48} V. V. Vien, H. N. Long and A.E. Cárcamo Hernández 
    \emph{Prog. Theor. Exp. Phys.} \textbf{2019} 113B04 (2019). 
\bibitem{LseesawS49} A.~E.~C\'{a}rcamo Hern\'{a}ndez, N.~A.~P\'{e}rez-Julve and Y.~Hidalgo Vel\'{a}squez 
    \emph{Phys. Rev. D} \textbf{100} 095025 (2019). 

\bibitem{scalarpoten1} V. V. Vien, H. N. Long and A. E. Cárcamo Hernández 
    \emph{Eur. Phys. J. C} \textbf{80} 725 (2020). 
\bibitem{scalarpoten2} V. V. Vien 
\emph{J. Phys. G: Nucl. Part. Phys.} \textbf{47}  055007 (2020). 
\bibitem{scalarpoten3} V. V. Vien 
\emph{Nucl. Phys. B} \textbf{956}  115015 (2020). 
\bibitem{scalarpoten4} V. V. Vien 
\emph{Mod. Phys. Lett. A} \textbf{35}  2050223 (2020). 



\bibitem{br1} L.~Lavoura, Eur. Phys. J. {\bf C29} 191 (2003).
\bibitem{br2} M.~D.~Campos, A.~E.~C\'{a}rcamo Hern\'{a}ndez, H.~P\"{a}s and E.~Schumacher, Phys.\ Rev.\ D {\bf 91}  116011 (2015).
\bibitem{br3} L.~T.~Hue, L.~D.~Ninh, T.~T.~Thuc and N.~T.~T.~Dat, Eur.\ Phys.\ J.\ C {\bf 78}  128 (2018).
\bibitem{br4} M.~Lindner, M.~Platscher and F.~S.~Queiroz, Phys.\ Rept.\  {\bf 731}, 1 (2018).

\bibitem{PDG2020} P. A. Zyla \emph{et al.} (Particle Data Group) 
\emph{Prog. Theor. Exp. Phys.} \textbf{2020} 083C01 (2020). 



\bibitem{Pontecorvo1} B. Pontecorvo 
\emph{Zh. Eksp. Teor. Fiz.} \textbf{33} 549 (1957).
\bibitem{Pontecorvo2} B. Pontecorvo 
\emph{Zh. Eksp. Teor. Fiz.} \textbf{34}  247 (1958).

\bibitem{Jcp} P. I. Krastev and S. T. Petcov 
\emph{Phys. Lett. B} \textbf{205} 84 (1988). 
\bibitem{Maki} Z. Maki, M. Nakagawa and S. Sakata 
\emph{Prog. Theor. Phys.} \textbf{28} 870 (1962). 
\bibitem{Rodejohann} W. Rodejohann 
\emph{Phys. Rev. D} \textbf{69} 033005 (2004). 

\bibitem{Jarlskog1} C. Jarlskog 
\emph{Phys. Rev. Lett.} \textbf{55} 1039 (1985). 
\bibitem{Jarlskog2} D. -d. Wu 
\emph{Phys. Rev. D} \textbf{33} 860 (1986). 
\bibitem{Jarlskog3} O. W. Greenberg 
\emph{Phys. Rev. D} \textbf{32} 1841 (1985). 




\bibitem{Capozzi20} F. Capozzi, E. D. Valentino, E. Lisi, A. Marrone, A. Melchiorri and A. Palazzo 
    \emph{Phys. Rev. D} \textbf{101} 116013 (2020).
\bibitem{nuboundAghanim20} N. Aghanim \emph{et al.} (Planck Collaboration) 
\emph{Astron. Astrophys.}  \textbf{641} A6(2020). 
\bibitem{nuboundShadab21} Shadab Alam \emph{et al.} 
\emph{Phys. Rev. D} \textbf{103}  083533 (2021). 


\bibitem{betdecay2} M. Mitra, G. Senjanovic and F. Vissani 
\emph{Nucl. Phys. B} \textbf{856} 26 (2012).
\bibitem{betdecay4} W. Rodejohann 
\emph{J. Phys. G} \textbf{39} 124008 (2012). 
\bibitem{betdecay5} J. D. Vergados, H. Ejiri and F. Simkovic 
\emph{Rep. Prog. Phys.} \textbf{75} 106301 (2012). 


\bibitem{EsfahaniP8} A. A. Esfahani \emph{et al.} 
    \emph{J. Phys. G} \textbf{44} 054004 (2017). 
\bibitem{Esfahani21} A. A. Esfahani \emph{et al.} 
    \emph{Phys. Rev. C} \textbf{103}  065501 (2021). 
\bibitem{Aker22n} M. Aker \emph{et al.} (The KATRIN Collaboration) 
    \emph{Nat. Phys.} \textbf{18} 160 (2022). 
\end{thebibliography}
\end{document}